\documentclass[%
 aip,
 jcp,
 preprint,
 amsmath,amssymb,
 twocols,
]{revtex4-1}

\usepackage[USenglish]{babel}
\usepackage{graphicx}
\usepackage{dcolumn}
\usepackage{bm}
\usepackage{color}
\usepackage{amsmath, amssymb, amsthm}
\usepackage{mathtools}
\usepackage{courier}
\usepackage{graphicx,color}
\usepackage{tabularx}
\usepackage[normalem]{ulem}
\usepackage[colorlinks=true,linkcolor=blue,citecolor=blue,urlcolor=blue]{hyperref}

\begin{document}


\title{From Classical to Quantum and Back: Hamiltonian Adaptive Resolution Path Integral, Ring Polymer, and Centroid Molecular Dynamics}

\author{Karsten Kreis}
\affiliation{Max-Planck-Institut f\"ur Polymerforschung, Ackermannweg 10, 55128 Mainz, Germany}
\affiliation{Graduate School Material Science in Mainz, Staudingerweg 9, 55128 Mainz, Germany}
\author{Kurt Kremer}
\affiliation{Max-Planck-Institut f\"ur Polymerforschung, Ackermannweg 10, 55128 Mainz, Germany}
\author{Raffaello Potestio}
\email{potestio@mpip-mainz.mpg.de}
\affiliation{Max-Planck-Institut f\"ur Polymerforschung, Ackermannweg 10, 55128 Mainz, Germany}
\author{Mark E. Tuckerman}
\email{mark.tuckerman@nyu.edu}
\affiliation{Department of Chemistry, New York University (NYU), New York, NY 10003, USA}
\affiliation{Courant Institute of Mathematical Sciences, NYU, New York, NY 10012, USA}
\affiliation{NYU-East China Normal University Center for Computational Chemistry at NYU Shanghai, Shanghai 200062, China}

\date{\today}

\begin{abstract}
Path integral-based simulation methodologies play a crucial role for the investigation of nuclear quantum effects by means of computer simulations. However, these techniques are significantly more demanding than corresponding classical simulations. To reduce this numerical effort, we recently proposed a method, based on a rigorous Hamiltonian formulation, which restricts the quantum modeling to a small but relevant spatial region within a larger reservoir where particles are treated classically. In this work, we extend this idea and show how it can be implemented along with state-of-the-art path integral simulation techniques, such as ring polymer and centroid molecular dynamics, which allow the approximate calculation of both quantum statistical and quantum dynamical properties. To this end, we derive a new integration algorithm which also makes use of multiple time-stepping. The scheme is validated via adaptive classical--path-integral simulations of liquid water. Potential applications of the proposed multiresolution method are diverse and include efficient quantum simulations of interfaces as well as complex biomolecular systems such as membranes and proteins.    
\end{abstract}


\maketitle

\section{Introduction} \label{introduction_9}
Quantum delocalization of light atomic nuclei plays an important role in many soft matter systems, ranging from low temperature helium or hydrogen \cite{Scharf1993a,Ceperley1995b,Azuah1997,Nozieres1999,Lindenau2006} to complex biological systems at room temperature. Examples include proton transfer in biomolecules and membranes \cite{Lowdin1963,Rein1964,Haines2001,Jeuken2007,Perez2010,Smirnov2011,Jacquemin2014}, thermodynamics of ice \cite{Pamuk2012}, water adlayers on catalysts \cite{Li2010a,Nagata2012}, aqueous proton and hydroxide transport \cite{Marx1999,Schmitt1999,Tuckerman2002,Wu2008,Berkelbach2009,Marx2010}, and even the structure and dynamics of bulk water \cite{Morrone2008,Paesani2010,Ceriotti2013,Fritsch2014,Ceriotti2016}.

In computer simulations, nuclear quantum effects are typically modeled using Feynman's path integral (PI) formulation of quantum statistical mechanics \cite{Feynman1965,Ceperley1995b,TuckermanBook}. The atomic nuclei are mapped onto classical ring polymers, whose beads correspond to the imaginary time slices of the PI. Based on this approach, various techniques have been developed to compute approximate quantum mechanical properties. Path integral molecular dynamics (PIMD) \cite{Chandler1981,Parrinello1984,Hall1984,Berne1986,Tuckerman1993,Martyna1999,TuckermanBook} and path integral Monte Carlo (PIMC) \cite{Pollock1984,Tuckerman1993,TuckermanBook} directly sample the Hamiltonian obtained after path integral quantization and can be employed to calculate time-independent quantum statistical properties. Centroid molecular dynamics (CMD) \cite{Cao1994d,Cao1994c,Cao1994e,Cao1994,Cao1994b,Cao1996,Martyna1996,Cao1996a,Jang1999,Hone2006,Perez2009,Polyakov2010}, which follows the artificial dynamics of the ring-polymer centroids, and ring polymer molecular dynamics (RPMD) \cite{Craig2004,Braams2006,Hone2006,Perez2009,Habershon2013,Rossi2014}, which is based on the evolution of the individual PI beads, additionally enable the calculation of approximate quantum dynamical properties.

However, PI-based methods are significantly more expensive than corresponding classical simulations. To overcome this, different techniques have been proposed. For example, ring polymer contraction (RPC) \cite{Markland2008,Markland2008a,Marsalek2016,John2016} makes use of the fact that long-ranged and non-bonded interactions typically do not need to be evaluated on as many PI beads as bonded interactions. A related technique is the mixed time slicing scheme \cite{Steele2011}, in which different particles are described with a different number of imaginary time slices. Other approaches include higher-order Trotter factorization \cite{Takahashi1984,Jang2001,Suzuki2010,Perez2011,Poltavsky2016} and advanced thermostating procedures based on generalized Langevin equations (GLE) \cite{Ceriotti2009,Ceriotti2010,Ceriotti2011,Ceriotti2012}. Additionally, multiple time-stepping (MTS) techniques are frequently employed in PI simulations to decouple the computation of the expensive but slowly varying non-bonded forces and the high frequency internal motion of the ring polymers \cite{Tuckerman1992,Tuckerman1993,Martyna1999}.

Most of these methods correspond to a modification of the path integral computation itself. A different approach is provided by adaptive resolution methods, which restrict the PI description to a small subregion within the simulation box and couple it with a classical model. The available computational resources can then be concentrated on the quantum (QM) subregion leading to an overall speedup compared with full QM simulations. This strategy is useful when only a small part of the overall large system actually needs to be described taking into account quantum delocalization effects, which might be the case, for example, in simulations of surfaces, membranes or the active site of a protein. One such method, based on the adaptive resolution simulation scheme (AdResS) \cite{Praprotnik2005,Praprotnik2008,Fritsch2012}, is the direct spatial interpolation of a classical force field with the PI-based forces obtained after quantization \cite{Poma2010,Poma2011,Potestio2012,Agarwal2015,Agarwal2016}. This method is, however, not compatible with an overall Hamiltonian description and, thus, inconsistent with the PI formalism \cite{DelleSite2007}. Nevertheless, it can, in principle, be used to simulate open quantum systems \cite{Wang2013a,Agarwal2015a}, for example.

Recently, we proposed a related multiresolution quantum--classical method that, instead of interpolating forces, directly changes the ``quantumness'', quantum character, or degree of quantum delocalization of the particles themselves \cite{Kreis2016}. In the QM region, the ring polymers are defined as usual, while in the classical region they collapse to point-like particles, thereby effectively behaving classically. When diffusing between the different regions, the particles change their resolution on the fly. Furthermore, the number of particles in the QM region is not fixed but is allowed to fluctuate. Hence, the scheme can, for example, be used to simulate a quantum grand canonical ensemble efficiently. The approach is derived in a rigorous fashion from the bottom up and is also compatible with a Hamiltonian description. When restricting the QM part to a small but relevant region in space, the scheme leads to a significant computational speedup. In the example of our previous paper, a liquid parahydrogen system, the calculation of the particle pair interactions was accelerated by a factor of $\approx10$. Furthermore, the approach can be combined with the previously mentioned methods for efficient PI simulations, such as RPC or GLE thermostating. Therefore, approaching the problem from a different perspective and reducing the number of PI-based interactions in the system, our method is complementary to techniques that make the PI computations themselves more efficient.

In our previous paper \cite{Kreis2016}, we proposed the general Hamiltonian adaptive quantum--classical scheme, performed a simple validation of the method using a Monte Carlo algorithm to sample the hybrid Hamiltonian, and demonstrated that the approach can speed up PI-based simulations. In this follow-up article, we show how the method can be extended to perform Hamiltonian multiresolution quantum--classical CMD and RPMD simulations. To this end, we derive an MTS integration protocol suited for the proposed methodology and validate the method by adaptive quantum--classical simulations of liquid water.

Our scheme enables efficient simulations of complex systems by locally taking into account QM delocalization effects. This can be useful, for example, for interface systems and in simulations of biological objects such as membranes or proteins. Additionally, as previously noted, it allows an efficient implementation of the QM grand canonical ensemble and can, in principle, also be combined with quantum mechanics/molecular mechanics (QM/MM) approaches, in particular those which are based on a similar Hamiltonian interpolation scheme \cite{Boereboom2016}.

The paper is organized as follows: In section \ref{methodology1_9}, we review the adaptive quantum--classical scheme proposed in our previous work, and in section \ref{methodology2_9} we present its implementation within the PIMD framework. In section \ref{methodology3_9}, we discuss how to use the methodology to calculate approximate quantum dynamical quantities in the context of adaptive RPMD and CMD simulations. We describe the details of the simulations we performed for validation in section \ref{simulations_9} and the results are discussed in section \ref{results_9}. In section \ref{conclusions_9}, we summarize the article and conclude.

\section{Quantum--classical path integrals} \label{methodology1_9}
In quantum statistical mechanics, the partition function of a system of $N$ interacting particles with indices $\alpha$, momenta $\hat{\mathbf{p}}_\alpha$, masses $m_\alpha$, kinetic energy $\hat{\mathcal{K}}=\sum_{\alpha=1}^N\hat{\mathbf{p}}_\alpha^2/2m_\alpha$ and potential energy $\hat{\mathcal{V}}$ is $Q = \textmd{Tr}[\exp{(-\beta \hat{\mathcal{H})}}]$ with the inverse temperature $\beta = 1/k_b T$ and the Hamiltonian $\hat{\mathcal{H}} = \hat{\mathcal{K}} + \hat{\mathcal{V}}$. When performing PI quantization using $P$ imaginary time slices (Trotter number $P$), the kinetic energy term gives rise to a configurational energy that is equivalent to the one of a classical ring polymer with $P$ beads, which are coupled via harmonic springs (for a detailed derivation see, for example, Tuckerman \cite{TuckermanBook}). This mapping from a quantum particle onto a classical polymer ring is exact in the limit $P\rightarrow\infty$. In practice, however, well converged results can be obtained for finite values of $P$, which typically range from 16 to 48 beads for standard PI simulations aimed at treating nuclear quantum effects under ambient conditions\cite{Perez2010,Tuckerman2001,Habershon2009,Fritsch2014,Wang2014a,Wilkins2015,Miller2005}.

The strength of the spring constants between the beads of the ring polymers is $m\omega^2_P$ with $\omega_P = \sqrt{P}/\beta\hbar$. It is proportional to the temperature as well as the particles' masses. In other words, the ring polymers are more collapsed the higher the temperature and the heavier the particles. The extension of the ring polymers can be interpreted as a measure of the ``quantumness'' or quantum character of the QM particles, with classical behavior corresponding to fully collapsed and therefore localized ring polymers.

\begin{figure}[ht!]
  \centering
\includegraphics[clip,width=0.95\columnwidth,keepaspectratio]{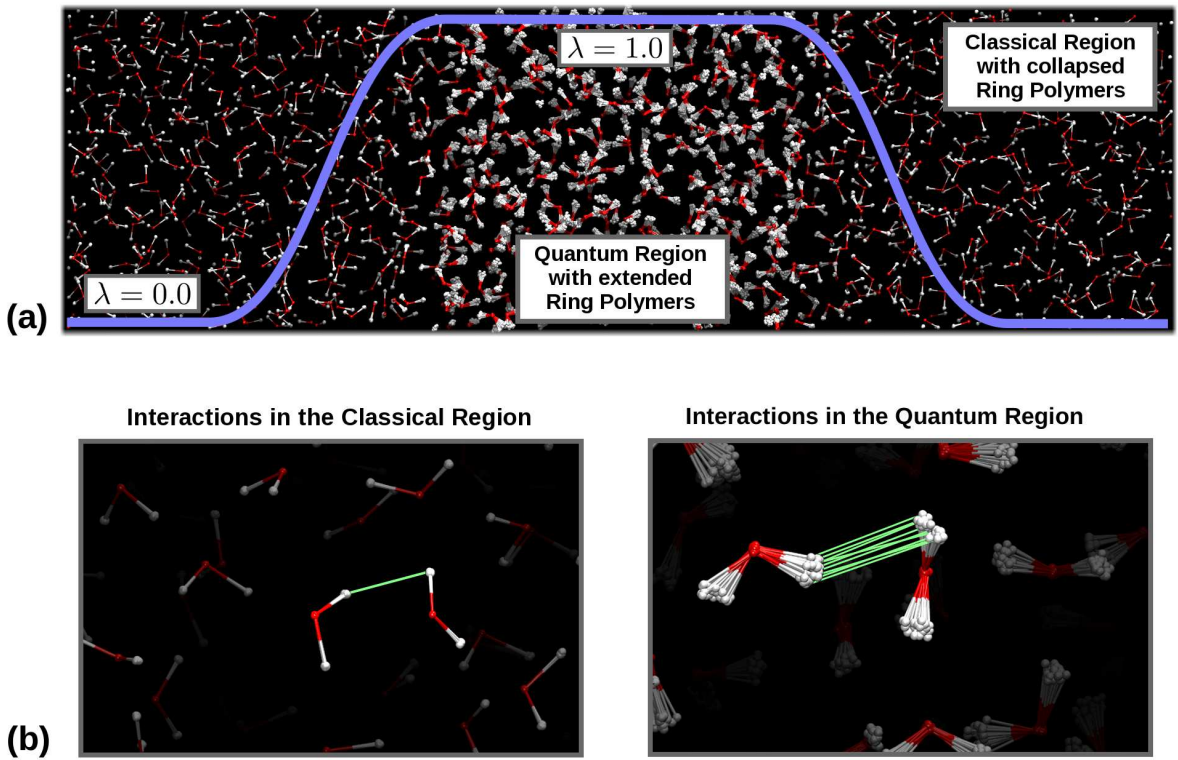}
\caption[Snapshots from the quantum--classical adaptive resolution path integral molecular dynamics simulations of liquid water.]{Simulation snapshots from the adaptive quantum--classical simulations. (a) A complete box of quantum--classical water: The blue line shows the resolution function $\lambda$ switching smoothly from $1$ in the QM region to $0$ in the CL region. In the QM region, the ring polymers, which correspond to the water atoms, are extended, modeling their quantum mechanical delocalization. In the CL region, they are collapsed to points. (b) Interacting atoms in the CL and QM regions: In the QM region, interactions between different atoms are given as the average over the $P$ time-slices. In the CL region, in contrast, only a single calculation is required for the interaction between a pair of atoms due to the point-like structure of the particles. This alleviates the numerical effort and reproduces normal classical computational efficiency in the CL subsystem.} 
\label{fig1_9}
\end{figure}

The previously proposed method for quantum--classical adaptive resolution simulations \cite{Kreis2016} is based on the following observation: In a PI-based formulation of quantum statistical mechanics the only role of a particle's mass is to determine the spring constant. It determines the extent to which a particle exhibits its quantum character. The scheme is as follows: For each particle $\alpha$ we define a resolution parameter $\lambda_\alpha=\lambda(\hat{\mathbf{r}}_\alpha)$ that is a function of the particle's position $\hat{\mathbf{r}}_\alpha$. It smoothly changes from $1$ in a spatially predefined QM region to $0$ in a classical (CL) region via an intermediate hybrid (HY) transition region (see Fig. \ref{fig1_9}).
Based on this resolution function, we then define a variable mass of particle $\alpha$ as $m_\alpha \rightarrow \mu_\alpha(\lambda_\alpha) = \lambda_\alpha m_\alpha + (1-\lambda_\alpha) M_\alpha$. Therefore, in the QM region $\mu_\alpha(1)=m_\alpha$ where $m_\alpha$ is the real mass of the particles while in the CL region $\mu_\alpha(0)=M_\alpha \gg m_\alpha$. The mass $M_\alpha$ must be chosen large enough that the particles with $\mu_\alpha(0)=M_\alpha$ behave essentially classically (in our previous work, we used $M_\alpha = 100\,m_\alpha$). In this way, particles in the QM region exhibit proper QM behavior, while in the CL region the ring polymers are forced to collapse to nearly point-like particles and behave classically.
Note, also, that the mass $M_\alpha$ is not employed, in the MC framework just discussed, to evolve a time-dependent dynamics of the system, which on the other hand is the object of the present study.

In addition to variable masses, we also use the Hamiltonian adaptive resolution simulation (H-AdResS) formalism \cite{Potestio2013,Potestio2013a,Kreis2014,Espanol2015}, which can be employed to couple different force fields via interpolation of potential energies. A classical H-AdResS Hamiltonian $H$ of a system of $N$ interacting molecules reads
\begin{eqnarray}\label{hadress_H}
H = \mathcal K + \sum_{\alpha=1}^N \left[{\lambda_\alpha} {V^{1}_\alpha} + {(1 - \lambda_\alpha)} {V^{0}_\alpha} + V^{\textrm{int}}_{\alpha} - \Delta H(\textbf{R}_\alpha) \right]
\end{eqnarray}
where $\mathcal K$ is the kinetic energy, $\alpha$ indexes the $N$ particles, and $\lambda_\alpha = \lambda(\textbf{R}_\alpha)$ is the previously defined resolution function (when assigning single resolution values $\lambda_\alpha$ to whole molecules one typically uses the molecular center of mass $\textbf{R}_\alpha$ as reference coordinate to determine $\lambda_\alpha$). The single-particle potentials $V^{Res}_\alpha$ (with $Res = 0, 1$) are the sums of all intermolecular potentials acting on particle $\alpha$, properly normalized so that double counting is avoided. The term $V^{\textrm{int}}_{\alpha}$ represents all intramolecular interactions, such as bond and angle potentials, which are not subject to interpolation. The term $\Delta H$, referred to as the Free Energy Compensation (FEC) \cite{Potestio2013,Potestio2013a}, is an external field acting in the HY transition region to eliminate the density imbalance that naturally occurs in such dual-resolution systems. Different models of the same physical system exhibit a free energy difference that needs to be neutralized in order to enforce identical thermodynamical and/or structural properties (e.g. density) everywhere in the simulation domain. The FEC levels off these free energy imbalances. H-AdResS has been used mainly to couple atomistic and coarse-grained two-body force fields \cite{Potestio2013,Potestio2013a,Kreis2015,Kreis2016b}. In general, however, the potentials can refer to any non-bonded interaction $V^{Res}_\alpha$.

We have combined the mass-based quantum--classical interpolation with the H-AdResS scheme in order to enable use of different force fields within the QM and the CL regions. This can be advantageous, for example, when the CL region's only role is that of a particle reservoir, in which case a very simple force field can be used in the CL region. On the other hand, when simulating a protein, for example, and only a small part of it needs to treat nuclear quantum effects explicitly, one would probably resort to the same force-field everywhere in the system and just ``add'' the nuclear quantum effects in the relevant region with the present quantum--classical multiresolution scheme.

Combining the quantum--classical mass interpolation and the H-AdResS scheme, we can write the Hamiltonian operator of $N$ interacting Boltzmann particles in three dimensions as \cite{Kreis2016}
\begin{equation}\label{eq:H1_9}
\hat{\mathcal{H}} = \sum_{\alpha=1}^N\left(\frac{1}{2}\hat{\mathbf{p}}_\alpha \mu^{-1}(\hat{\mathbf{r}}_\alpha)\hat{\mathbf{p}}_\alpha  + V^{\textrm{H-AdResS}}(\hat{\mathbf{r}}_\alpha)\right)
\end{equation}
where $\mu^{-1}(\hat{\mathbf{r}}_\alpha)$ is the inverse mass operator. The potential energy term $V^{\textrm{H-AdResS}}(\hat{\mathbf{r}}_\alpha)$ corresponds to the interpolated H-AdResS potential energy, i.e. the term within the sum in Eq. \ref{hadress_H}. We have shown that PI quantization then leads to the following approximate expression for the partition function \cite{Kreis2016} (the exact nature of the approximation will be discussed shortly):
\begin{equation}\label{eq:H12_9}
Q=\lim\limits_{P\to\infty}\left[ \prod_{k=1}^P\prod_{\alpha=1}^N 
\int\text{d}{\mathbf{r}}_{\alpha,k}\left( \frac{\mu_{\alpha,k}P}{2\pi\beta\hbar^2} \right)^{\frac{3}{2}}\right]
e^{-\beta V_P^\mu}
\end{equation}
with 
\begin{equation}\label{vp:2_9}
\begin{split}
V_P^\mu &= \sum_{k = 1}^P\ \sum_{\alpha = 1}^N \left\{ \frac{\mu_{\alpha,k}\ \omega_P^2}{2} |{\mathbf{r}}_{\alpha,k} - \mathbf{r}_{\alpha,k+1}|^2
\right. \\
&+\left. \frac{1}{P} \left [\lambda_{\alpha,k} V^{\textrm{QM}}_{\alpha,k} + (1 - \lambda_{\alpha,k}) V^{\textrm{CL}}_{\alpha,k}  + V^{\textrm{int}}_{\alpha,k} - \Delta H(\mathbf{r}_{\alpha,k}) \right] \right\}
\end{split}
\end{equation}
where $\alpha$ indexes the different particles and $k$ the individual Trotter beads for each of them. A resolution value $\lambda_{\alpha,k}$ is associated with each bead. We have renamed $V^{0,1}_{\alpha,k}$ to $V^{\textrm{CL},\textrm{QM}}_{\alpha,k}$ to emphasize that $V^{\textrm{CL}}_{\alpha,k}$ ($V^{\textrm{QM}}_{\alpha,k}$) is the intermolecular potential acting in the CL (QM) region. Note that the normalization term in Eq. \ref{eq:H12_9} depends on the position of the particles via $\mu_{\alpha,k}$. To obtain a constant normalization factor, one can transform this position dependent term to a potential energy in $V_P^\mu$, as done in our previous article \cite{Kreis2016}, and then treat it as a constant field in the hybrid region. In this work, however, we will deal with it in a different way, which we will discuss in detail later.

The above expression (Eqs. \ref{eq:H12_9} and \ref{vp:2_9}) is consistent with a rigorous PI quantization if
\begin{equation}\label{eq:H11_9}
\left| \frac{d\mu(x)}{dx} \right| \ll \frac{2\mu(x)}{\Lambda_\mu(x)}
\end{equation}
with $\Lambda_\mu(x) \equiv \sqrt{\beta\hbar^2/(P\mu(x))}$. A derivation of this condition was presented in Ref.~\onlinecite{Kreis2016}.  This criterion indicates that the interpolation between the QM and the CL parts of the system must be sufficiently smooth such that the mass difference between two neighboring beads on a ring polymer in the HY region is negligible. This requirement can always be satisfied by choosing a sufficiently large HY region. Note, however, that even if Eq. \ref{eq:H11_9} is not fulfilled, Eqs. \ref{eq:H12_9} and \ref{vp:2_9} correspond to a well-defined quantum--classical simulation protocol.

The ring polymers described by the energy function $V_P^\mu$ (Eq. \ref{vp:2_9}) are expanded in the region where the mass is small, and collapse to nearly classical point-like particles in the large-mass region. Therefore, in the CL region the interactions between different ring polymers do not need to be computed as an average over the $P$ bead pairs as done in the QM region. Instead, due to their nearly point-like structure, one can use only the centroid with negligible error (see Fig. \ref{fig1_9}). In this manner, classical computational efficiency is regained in the CL region.

\section{Quantum--classical path integral molecular dynamics} \label{methodology2_9}
In our previous paper \cite{Kreis2016}, we validated the scheme introduced above via simulations of liquid parahydrogen using a basic Monte Carlo algorithm to directly sample the phase space defined by $V_P^\mu$. A more state-of-the-art approach to the numerical evaluation of PIs is provided by PIMD, in which a thermostated dynamics is generated in phase space to sample the quantum canonical ensemble \cite{Chandler1981,Parrinello1984,Hall1984,Berne1986,Tuckerman1993,Martyna1999,TuckermanBook}. PIMD is more easily parallelizable compared to Monte Carlo methods and therefore significantly more efficient for typical simulation setups and on multicore computer architectures. Moreover, PIMD is the method of choice for {\it ab initio} path integrals \cite{AIPIMD1,AIPIMD2}.  In the following, we show how the proposed quantum--classical multiresolution method can be implemented in PIMD as well as CMD and RPMD.

\subsection{Evaluation of the adaptive mass and the resolution function on the centroids} \label{methodology2_1_9}
In order to decouple the modes of the cyclic ring polymers from each other, PIMD is typically performed using staging variables \cite{Pollock1984,Tuckerman1993} or, more popularly, normal modes \cite{Tuckerman1993,Cao1994}. In our case, we cannot transform smoothly into normal mode space, because the beads of the individual ring polymers have different masses $\mu_{\alpha,k}$ within the HY region. For typical systems like liquid water at room temperature, however, this mass difference is small as the extension of the ring polymers, measured by the root-mean-square radius of gyration $r^g$, is short, even in the QM region. This suggests that we  associate a single resolution value $\lambda_\alpha$ and a single adaptive mass value $\mu_\alpha$ with each atomic or molecular particle $\alpha$ instead of with every single bead $k$. $\lambda_\alpha$ and $\mu_\alpha$ can then be determined using the ring polymer's centroid positions.

Although this corresponds only to a minor modification in Eq. \ref{vp:2_9}, we can ask to what extent the configurational energy $V_P^\mu$ is then still compatible with formal PI quantization. To this end, we first consider the adaptive mass $\mu_{\alpha,k}$ of an individual Trotter bead within the HY region which we can approximate as
\begin{equation}\label{EQ:4}
\mu_{\alpha,k} \approx \mu^c_{\alpha}+\delta x_{\alpha,k}\frac{d\mu(x^c_\alpha)}{dx},
\end{equation}
where $x^c_\alpha$ is the centroid coordinate along the direction of resolution change of the ring polymer $\alpha$. For a setup where the resolution changes along the x-direction, this would be the centroid's $x$-coordinate and for a system where the QM region is spherical and the resolution changes radially, this would be the radial distance from the center. $\mu^c_{\alpha}$ is the mass function evaluated at the centroid $x^c_\alpha$ and $\delta x_{\alpha,k}$ is the distance along the direction of resolution change between the $k$-th bead of ring $\alpha$ and its centroid $x^c_\alpha$. We have $\delta x_{\alpha,k}\lesssim r^g(x^c_\alpha)$, where $r^g(x) = \sqrt{\beta\hbar^2/(4\mu(x))} \cdot\sqrt{1-1/P^2}\approx \sqrt{\beta\hbar^2/(4\mu(x))}$ is the radius of gyration of a free ring with mass $\mu(x)$ for large $P$. Hence, we can approximate $\mu_{\alpha,k} \approx \mu^c_{\alpha}$ if 
\begin{equation}\label{EQ:5}
\left|\frac{d\mu(x^c_\alpha)}{dx}\right| \ll \frac{\mu^c_{\alpha}}{r^g(x^c_\alpha)},
\end{equation}
or simply 
\begin{equation}\label{EQ:6}
\left|\frac{d\mu(x)}{dx}\right| \ll \frac{\mu(x)}{r^g(x)}.
\end{equation}
for general $x$. Here, we used the free ring-polymer radius of gyration. However, for the ring polymers in typical systems the average radius of gyration differs only slightly from the free-particle radius.

Eq. \ref{EQ:6} is a slightly stronger criterion than that in Eq. \ref{eq:H11_9}. This makes sense, since Eq. \ref{eq:H11_9} essentially provides the condition under which the mass can be considered as constant between two neighboring beads, while the new criterion, Eq. \ref{EQ:6}, gives the condition for treating the mass as constant over an entire ring polymer. For liquid water at room temperature, Eq. \ref{EQ:6} is satisfied by a hybrid region wider than $\approx 1\,\text{nm}$. An even smaller hybrid region would not be desirable anyway, since the interaction cutoffs of typical interaction potentials are also of the order $\approx 1\,\text{nm}$.

Next, we consider the resolution $\lambda$ itself, which we also seek to treat as constant over a whole ring polymer so that it can be approximated as a function of the centroid position only. On the one hand, $\lambda$ varies between 0 and 1 and changes most steeply in the center of the HY region (see Fig. \ref{fig1_9}). On the other hand, an upper bound on the extension of the ring polymers is provided by the radius of gyration of the ring polymers in the QM region, which will be denoted as $r_{\text{QM}}^g$. Therefore, if the change in $\lambda$, corresponding to its gradient in the center of the HY region, over a distance $r_{\text{QM}}^g$ is much smaller than $\approx1$, we can approximate $\lambda_{\alpha,k} \approx \lambda^c_{\alpha}$ everywhere (as for the adaptive mass before, $\lambda^c_{\alpha}$ denotes the resolution function of ring $\alpha$ evaluated at its centroid $x^c_\alpha$). This corresponds to
\begin{equation}\label{EQ:7}
\left|\left(\frac{d\lambda(x)}{dx}\bigg|_{x_=d_\text{HY}/2}\right)\right| r^g_{\text{QM}} \ll 1,
\end{equation}
where $d\lambda(x)/dx|_{x=d_\text{HY}/2}$ denotes the gradient of $\lambda$ in the center of the HY region and $d_\text{HY}$ the width of the HY region. The criterion in Eq. \ref{EQ:7} can be easily fulfilled for typical systems such as water at room temperature, with a HY region of width $d_\text{HY}\approx 1\,\text{nm}$.

To summarize, in the following we will treat both the mass and the resolution as a constant over entire ring polymers and write, for simplicity, $\mu_{\alpha,k}(x)\rightarrow\mu_\alpha(x)$ and $\lambda_{\alpha,k}(x)\rightarrow\lambda_{\alpha}(x)$, where we assume that the mass and the resolution functions for an atom $\alpha$ have been evaluated using its centroid coordinate $x^c_\alpha$ and that the resulting parameters have been assigned to all beads belonging to ring $\alpha$. Then we can write the partition function as
\begin{equation}\label{EQ:9}
Q=\lim\limits_{P\to\infty}\left[ \prod_{k=1}^P\prod_{\alpha=1}^N \int\text{d}{\mathbf{r}}_{\alpha,k}\left( \frac{\mu_{\alpha}P}{2\pi\beta\hbar^2} \right)^{\frac{3}{2}}\right]
e^{-\beta \tilde{V}_P^\mu}
\end{equation}
with 
\begin{eqnarray}\label{EQ:10}
\tilde{V}_P^\mu &=& \sum_{k = 1}^P\ \sum_{\alpha = 1}^N \left\{ \frac{\mu_{\alpha}\ \omega_P^2}{2} |{\mathbf{r}}_{\alpha,k} - {\mathbf{r}}_{\alpha,k+1}|^2
\right. \nonumber \\
&+&\left. \frac{1}{P} \left [\lambda_{\alpha} V^{\textrm{QM}}_{\alpha,k} + (1 - \lambda_{\alpha}) V^{\textrm{CL}}_{\alpha,k}  + V^{\textrm{int}}_{\alpha,k} - \Delta H(\mathbf{r}^c_{\alpha}) \right] \right\},
\end{eqnarray}
where $\mathbf{r}^c_{\alpha}$ denotes the centroid of ring polymer $\alpha$.
Note that the FEC is now also applied at the single-atom, i.e., centroid level. The criteria in Eqs. \ref{EQ:6} and \ref{EQ:7} quantify the degree to which this partition function is still compatible with a formal, bottom-up PI quantization. The inequalities can always be fulfilled by choosing a sufficiently wide hybrid region. However, even if they are not met, the final partition function, Eqs. \ref{EQ:9} and \ref{EQ:10}, still represents a well-defined Hamiltonian multiresolution quantum--classical simulation scheme.

\subsection{Introducing normal modes} \label{methodology2_2_9}
Now that the different beads of each ring polymer $\alpha$ all have the same adaptive mass $\mu_\alpha$, we can proceed with transforming the Cartesian coordinates into normal modes $\mathbf{u}_{\alpha,k}$ via
\begin{equation} \label{eq:modetrafo}
\mathbf{u}_{\alpha,k} = \sum^P_{j=1} \mathbf{r}_{\alpha,j} C_{jk},
\end{equation}
where, for even $P$, the orthogonal transformation matrix is \cite{Ceriotti2010}
\begin{equation}
C_{jk} =
\left\{
	\begin{array}{ll}
		\sqrt{1/P}\,  & \mbox{if } k=1 \\
		\sqrt{2/P}\,\textrm{cos}(2\pi jk/P)  & \mbox{if } 2 \leq k \leq P/2 \\
        \sqrt{1/P}\,(-1)^j  & \mbox{if } k=P/2+1 \\
        \sqrt{2/P}\,\textrm{sin}(2\pi jk/P) & \mbox{if } P/2+2 \leq k \leq P
	\end{array}
\right.
\end{equation}
such that for a given ring polymer at position $x$:
\begin{equation} 
\mu_{\alpha}(x)\sum_{k=1}^P|{\mathbf{r}}_{\alpha,k} - {\mathbf{r}}_{\alpha,k+1}|^2 = \mu_{\alpha}(x) \sum_{l=1}^P \xi_k \mathbf{u}^2_{\alpha,k}
\end{equation}
with 
\begin{equation} 
\xi_k = 4\,\textrm{sin}^2\left(\frac{(k-1)\pi}{P}\right).
\end{equation}
In normal mode representation, the centroid of a ring polymer $\alpha$ is given by the rescaled first mode coordinate, this is
\begin{equation}
\textbf{r}_\alpha^c = \frac{1}{P}\sum_{k=1}^P \textbf{r}_{\alpha,k}=\frac{1}{\sqrt{P}}\mathbf{u}_{\alpha,1}.
\end{equation}
Therefore, the adaptive mass $\mu_\alpha$ and the resolution function $\lambda_\alpha$, being evaluated at the centroids, are functions of the first mode only, i.e. $\mu_\alpha(\textbf{r}_\alpha^c)=\mu_\alpha(1/\sqrt{P}\,\mathbf{u}_{\alpha,1})$ and $\lambda_\alpha(\textbf{r}_\alpha^c)=\lambda_\alpha(1/\sqrt{P}\,\mathbf{u}_{\alpha,1})$. To lighten the notation, though, we will drop the $1/\sqrt{P}$ factor and write simply $\mu_\alpha=\mu_\alpha(\mathbf{u}_{\alpha,1})$ and $\lambda_\alpha=\lambda_\alpha(\mathbf{u}_{\alpha,1})$.

We then obtain the following partition function
\begin{equation}\label{EQ:11}
Q=\lim\limits_{P\to\infty}\left[ \prod_{k=1}^P\prod_{\alpha=1}^N  \int\text{d}{\mathbf{u}}_{\alpha,k}\left( \frac{\mu_{\alpha}(\mathbf{u}_{\alpha,1})P}{2\pi\beta\hbar^2} \right)^{\frac{3}{2}}\right]
e^{-\beta \tilde{V}_P^\mu}
\end{equation}
with 
\begin{equation}\label{EQ:12}
\begin{split}
\tilde{V}_P^\mu =& \sum_{k = 1}^P\ \sum_{\alpha = 1}^N \left\{ \frac{1}{2} \nu_{\alpha,k}(\mathbf{u}_{\alpha,1}) \omega_P^2 {\mathbf{u}}_{\alpha,k}^2
+ \frac{1}{P} \left [\lambda_{\alpha}(\mathbf{u}_{\alpha,1}) V^{\textrm{QM}}_{\alpha,k}(\mathbf{u}) + \right. \right.\\
&\left.\left. + \left(1 - \lambda_{\alpha}(\mathbf{u}_{\alpha,1})\right) V^{\textrm{CL}}_{\alpha,k}(\mathbf{u})  + V^{\textrm{int}}_{\alpha,k}(\mathbf{u}) - \Delta H(\mathbf{u}_{\alpha,1}) \right] \right\}
\end{split}
\end{equation}
and the rescaled adaptive mass $\nu_{\alpha,k}(\mathbf{u}_{\alpha,1})=\mu_{\alpha}(\mathbf{u}_{\alpha,1})\,\xi_k$. For the centroid, i.e. $k=1$, we have $\nu_{\alpha,k}(\mathbf{u}_{\alpha,1})=0$. In Eqs. \ref{EQ:11} and \ref{EQ:12}, we have explicitly indicated the dependencies of the different terms on the normal modes. The notation $\mathbf{u}$ without any indices is a shorthand for the compound set of all coordinates $\mathbf{u}_{\alpha,k}$.

\subsection{Introducing momenta} \label{methodology2_3_9}
In PIMD, one usually recasts the prefactor of the partition function as Gaussian integrals over a set of variables that can be interpreted as momenta conjugate to the coordinates $\mathbf{u}_{\alpha,k}$ \cite{TuckermanBook}. The energy term in the exponential can then be interpreted as a classical Hamiltonian and sampled via thermostated molecular dynamics. In our case, however, the prefactor is position dependent and, therefore, we have different options to proceed.

\textbf{(a) Constant kinetic masses}. As was done in our previous article \cite{Kreis2016}, we can write
\begin{equation} \label{EQ:12_2}
\left( \frac{\mu_{\alpha}(\mathbf{u}_{\alpha,1})P}{2\pi\beta\hbar^2} \right)^{\frac{3}{2}} = \left( \frac{\tilde{m} P}{2\pi\beta\hbar^2} \right)^{\frac{3}{2}} \exp\left\{-\beta\left(-\frac{3}{2\beta}\log\left(\frac{\mu_\alpha(\mathbf{u}_{\alpha,1})}{\tilde{m}}\right) \right)\right\}
\end{equation}
where we used an arbitrary mass $\tilde{m}$ as the reference mass scale. Then we can pull the term $-(3/2\beta)\,\log\left(\mu_\alpha/\tilde{m}\right)$ into $\tilde{V}_P^\mu$ as an external field in the HY region and use the now constant prefactor to introduce a set of momenta via rephrasing the prefactor as Gaussian integrals. This yields  
\begin{equation}\label{EQ:13}
Q=\lim\limits_{P\to\infty}\left[ \prod_{k=1}^P\prod_{\alpha=1}^N  \int\text{d}{\mathbf{u}}_{\alpha,k}\int\text{d}{\mathbf{p}}_{\alpha,k}\right]
e^{-\beta H_P^\textrm{ckm}}
\end{equation}
with the Hamiltonian
\begin{equation}\label{EQ:14}
\begin{split}
H_P^\textrm{ckm} =& \sum_{k = 1}^P\ \sum_{\alpha = 1}^N \left\{\frac{{\mathbf{p}}_{\alpha,k}^2}{2m'_{\alpha,k}} + \frac{1}{2} \nu_{\alpha,k} \omega_P^2 {\mathbf{u}}_{\alpha,k}^2  - \frac{3}{2P\beta} \log{\frac{\mu_{\alpha}({\mathbf{u}}_{\alpha,1})}{\tilde{m}}} + \right.\\
&\left. + \frac{1}{P} \left [\lambda_{\alpha}({\mathbf{u}}_{\alpha,1}) V^{\textrm{QM}}_{\alpha,k}({\mathbf{u}}) + \left(1 - \lambda_{\alpha}({\mathbf{u}}_{\alpha,1})\right) V^{\textrm{CL}}_{\alpha,k}({\mathbf{u}})  + V^{\textrm{int}}_{\alpha,k}({\mathbf{u}}) - \Delta H\left({\mathbf{u}}_{\alpha,1}\right) \right] \right\}.
\end{split}
\end{equation}
As the additional logarithmic term only acts as an external field in the HY region, it can be exactly removed via the FEC function $\Delta H$. We will denote the fictitious masses $m'_{\alpha,k}$ in the following as ``kinetic'' masses in contrast to $\nu_{\alpha,k}$, which we will refer to as spring masses. In principle, the set of $m'_{\alpha,k}$ can be chosen freely, as their rescaling does not affect thermodynamic averages \cite{TuckermanBook}.

$H_P^\textrm{ckm}$ in Eq. \ref{EQ:14} defines a classical Hamiltonian system composed of ring polymers representing the delocalized atoms. In the QM region where $\lambda_\alpha = 1$ and $\mu_\alpha=m$, the ring polymers are extended and the regular quantum mechanical behaviour is recovered. In the CL region where $\lambda_\alpha = 0$ and $\mu_\alpha=M$ the ring polymers are collapsed to essentially point-like particles, thereby reproducing classical mechanics. The Hamiltonian gives rise to regular equations of motion that can be integrated by a symplectic integrator such as the velocity Verlet algorithm \cite{TuckermanBook}, with the possibility of employing multiple time-stepping.

However, let us take a closer look at the different masses in the system. While the spring masses $\nu_{\alpha,k}$ change between the CL and QM subregions of the system, the kinetic masses $m'_{\alpha,k}$ do not. We choose $m'_{\alpha,k}=m_{\alpha}/P$ with $m_{\alpha}$ being the real mass of atom $\alpha$, since this corresponds to a realistic bead-wise approximate quantum dynamical behavior in the QM region similar to RPMD (in RPMD one usually chooses $m'_{\alpha,k}=m_{\alpha}$ without $1/P$, but rescales the potential energy terms by $P$ and runs the simulation at a $P$-times higher temperature \cite{Craig2004,Braams2006,Hone2006,Perez2009,Habershon2013,Rossi2014}. Here, the factor of $1/P$ in the mass is equivalent to this procedure, as we perform the simulations at the actual temperature and use a Hamiltonian, Eq. \ref{EQ:14}, without rescaling potential energies). Therefore, in the QM region, the modes oscillate with vibration frequencies $\omega_P\sqrt{\xi_kP}$. In the CL region, however, where $\nu_{\alpha,k}$ is significantly larger than in the QM subsystem the modes oscillate faster than in the QM region by a factor of $\sqrt{M_\alpha / m_\alpha}$. For the case of $M_\alpha=100\,m_\alpha$, this results in 10 times higher frequencies. This would require a 10 times smaller time step in the integration algorithm compared to a normal quantum simulation or compared to what would be required in the QM subregion. Although this poses no fundamental hurdle, it may slow down the simulations notably. 

\textbf{(b) Adaptive kinetic masses.}
The previous observation suggests an alternative approach: We can also directly recast the prefactor as a Gaussian integral, which includes the position dependent mass $\mu_\alpha$,
\begin{equation}\label{EQ:15}
\left( \frac{\mu_{\alpha}(\mathbf{u}_{\alpha,1})P}{2\pi\beta\hbar^2} \right)^{\frac{3}{2}} = \left(\frac{P^2}{4\pi^2\hbar^2}\right)^{\frac{3}{2}} \int d\boldsymbol{p}_{\alpha,k} \exp \left\{-\beta P\,\frac{\boldsymbol{p}_{\alpha,k}^2}{2\mu_{\alpha}(\mathbf{u}_{\alpha,1})} \right\}.
\end{equation}
In this way, we can introduce a kinetic energy term which has adaptive kinetic masses. This leads to the construction of a Hamiltonian in which both the spring and the kinetic masses vary in the same fashion, such that the modes oscillate with the same frequencies everywhere in the quantum--classical adaptive resolution setup.

Specifically, we propose the following: Overall, we have $N\times P$ prefactors of the form
\begin{equation}
\left( \frac{\mu_{\alpha}(\mathbf{u}_{\alpha,1})P}{2\pi\beta\hbar^2} \right)^{\frac{3}{2}},
\end{equation}
one for each atom and mode. For all higher modes with $k>1$, we transform the prefactors according to Eq. \ref{EQ:15} and introduce the momentum terms in the Hamiltonian with a variable mass in the denominator. The remaining $N$ prefactors are then treated via Eq. \ref{EQ:12_2}, and the kinetic masses for the centroid modes, which are not associated with springs since $\nu_{\alpha,1}=0$, are chosen constant. We then obtain
\begin{equation}\label{EQ:16}
Q=\lim\limits_{P\to\infty}\left[ \prod_{k=1}^P\prod_{\alpha=1}^N  \int\text{d}{\mathbf{u}}_{\alpha,k}\int\text{d}{\mathbf{p}}_{\alpha,k}\right]
e^{-\beta H_P^\textrm{akm}}
\end{equation}
with the Hamiltonian
\begin{equation}\label{EQ:17}
\begin{split}
H_P^\textrm{akm} =& \sum_{k = 1}^P\ \sum_{\alpha = 1}^N \left\{\frac{{\mathbf{p}}_{\alpha,k}^2}{2\nu'_{\alpha,k}({\mathbf{u}}_{\alpha,1})} + \frac{1}{2} \nu_{\alpha,k} \omega_P^2 {\mathbf{u}}_{\alpha,k}^2  - \frac{3}{2P\beta} \log{\frac{\mu_{\alpha}({\mathbf{u}}_{\alpha,1})}{\tilde{m}}} + \right.\\
&\left. + \frac{1}{P} \left [\lambda_{\alpha}({\mathbf{u}}_{\alpha,1}) V^{\textrm{QM}}_{\alpha,k}({\mathbf{u}}) + \left(1 - \lambda_{\alpha}({\mathbf{u}}_{\alpha,1})\right) V^{\textrm{CL}}_{\alpha,k}({\mathbf{u}})  + V^{\textrm{int}}_{\alpha,k}({\mathbf{u}}) - \Delta H\left({\mathbf{u}}_{\alpha,1}\right) \right] \right\},
\end{split}
\end{equation}
where $\nu'_{\alpha,k}$ is the kinetic mass of bead $k$ of atom $\alpha$. Note that a $1/P$ factor appears in front of the logarithmic term in $H_P^\textrm{akm}$ because the term still appears in the sum over all $P$, although we obtain the logarithmic term only for the centroid modes. Choosing appropriate prefactors, the parameters $\nu'_{\alpha,k}$ are
\begin{eqnarray} \label{kineticmasses1}
\nu'_{\alpha,1} = & m_\alpha/P, & \qquad k=1,\textrm{ centroid mode},\\ \label{kineticmasses2}
\nu'_{\alpha,k} = & \tilde{\nu}_{\alpha} = \mu_\alpha/P, & \qquad k>1,\textrm{ higher modes}.
\end{eqnarray}
Since the kinetic masses for the higher modes are all equal, we introduced a new abbreviation, $\tilde{\nu}_{\alpha}$, for them without the index $k$. We choose a factor $1/P$ to ensure that the approximate quantum dynamical time evolution of the centroids proceeds on the real timescale, that is, the same as in corresponding classical Newtonian dynamics. As already pointed out, this choice is equivalent to the temperature rescaling often done in ring polymer molecular dynamics \cite{Craig2004,Braams2006,Hone2006,Perez2009,Habershon2013,Rossi2014}, which we do not perform here. A further rescaling of the kinetic masses would be allowed when sampling only canonical averages \cite{TuckermanBook,Tuckerman1993}.

We can interpret the choice of the kinetic masses in the following way: While moving from the QM to the CL via the HY region, the spring constants of the higher modes become stronger and the ring polymers collapse. Simultaneously, however, these higher modes become heavier such that they do not vibrate faster in the CL region than in the QM region, despite the stiffer springs. Their oscillation frequencies are the same everywhere in the system such that their configurations can be sampled efficiently throughout the whole system with the same time step. The centroid modes do not undergo such oscillations, as they represent only the displacements of the ring polymers as a whole. Hence, their masses do not need to change across the transition from the QM to the CL part of the system and, therefore, they are chosen to be the real masses.

In the following we will refer to the constant kinetic mass (CKM) approach, defined by the Hamiltonian $H_P^\textrm{ckm}$ in Eq. \ref{EQ:14}, as the CKM approach and to the adaptive kinetic mass (AKM) scheme, defined by the Hamiltonian $H_P^\textrm{akm}$ in Eq. \ref{EQ:17}, as the AKM approach.

\subsection{Equations of motion} \label{methodology2_4_9}
We will not discuss the equations of motion obtained in the CKM approach, as they resemble a regular structure that can be integrated, for example, by a regular velocity Verlet algorithm \cite{TuckermanBook}. Instead, we focus on the AKM scheme, from which the CKM approach can be obtained as a special case.

In the following, we will assume that the logarithmic term in the Hamiltonian in Eq. \ref{EQ:17} has been exactly canceled by an appropriately chosen FEC function $\Delta H(\lambda)$ and therefore omit it to simplify the notation.

The equations of motion then read as follows:
\begin{equation}\label{EqMotion1}
\dot{\boldsymbol{u}}_{\alpha,1} = \frac{\boldsymbol{p}_{\alpha,1}}{\nu'_{\alpha,1}(\boldsymbol{u}_{\alpha,1})} = P\frac{\boldsymbol{p}_{\alpha,1}}{m_\alpha},\qquad k=1,\textrm{ centroid mode,}
\end{equation}
\begin{equation}\label{EqMotion2}
\dot{\boldsymbol{u}}_{\alpha,k} = \frac{\boldsymbol{p}_{\alpha,k}}{\tilde{\nu}_{\alpha}(\boldsymbol{u}_{\alpha,1})} = P\frac{\boldsymbol{p}_{\alpha,k}}{\mu_\alpha(\boldsymbol{u}_{\alpha,1})},\qquad k>1,\textrm{ higher modes,}
\end{equation}
\begin{equation}\label{EqMotion3}
\begin{split}
\dot{\boldsymbol{p}}_{\alpha,1} & = \mathbf{F}^\textrm{H-AdResS}_{\alpha,1}(\mathbf{u})\\
& - \frac{1}{P}\left[ \sum_{i=1}^P  \left\{ V^{\textrm{QM}}_{\alpha,i}(\boldsymbol{u}) - V^{\textrm{CL}}_{\alpha,i}(\boldsymbol{u}) \right\} \right] \boldsymbol{\nabla}_{\boldsymbol{u}_{\alpha,1}} \lambda(\boldsymbol{u}_{\alpha,1})\qquad\qquad(\mathbf{F}_1^\textrm{drift}) \\
& - \left[ \sum_{i=2}^P \frac{1}{2} \frac{(M_\alpha-m_\alpha)}{P\tilde{\nu}_{\alpha}(\boldsymbol{u}_{\alpha,1})^2} \boldsymbol{p}_{\alpha,i}^2 \right] \boldsymbol{\nabla}_{\boldsymbol{u}_{\alpha,1}} \lambda(\boldsymbol{u}_{\alpha,1})\qquad\qquad\qquad(\mathbf{F}_2^\textrm{drift}) \\
& - \left[ \sum_{i=2}^P \frac{1}{2} \xi_{i} (m_\alpha-M_\alpha) \omega_P^2 {\mathbf{u}}_{\alpha,i}^2 \right] \boldsymbol{\nabla}_{\boldsymbol{u}_{\alpha,1}} \lambda(\boldsymbol{u}_{\alpha,1})\qquad\qquad\qquad(\mathbf{F}_3^\textrm{drift}) \\
& + \boldsymbol{\nabla}_{\boldsymbol{u}_{\alpha,1}} \Delta H(\boldsymbol{u}_{\alpha,1}), \\
& \qquad\qquad\qquad\qquad\qquad\qquad\qquad\qquad\qquad k=1,\textrm{ centroid mode,}
\end{split}
\end{equation}
\begin{equation}\label{EqMotion4}
\dot{\boldsymbol{p}}_{\alpha,k} = -\nu_{\alpha,k}(\boldsymbol{u}_{\alpha,1})\omega_P^2{\mathbf{u}}_{\alpha,k}+\mathbf{F}^\textrm{H-AdResS}_{\alpha,k}(\mathbf{u}),\qquad k>1,\textrm{ higher modes,}
\end{equation}
where 
\begin{equation}\label{EqMotion5}
\mathbf{F}^\textrm{H-AdResS}_{\alpha,k}(\mathbf{u}) = -\frac{1}{P}\sum_{j = 1}^P\ \sum_{\beta = 1}^N \left[ \lambda_{\beta} \boldsymbol{\nabla}_{\boldsymbol{u}_{\alpha,k}} V^{\textrm{QM}}_{\beta,j}({\mathbf{u}}) + \left(1 - \lambda_{\beta}\right) \boldsymbol{\nabla}_{\boldsymbol{u}_{\alpha,k}} V^{\textrm{CL}}_{\beta,j}({\mathbf{u}})  + \boldsymbol{\nabla}_{\boldsymbol{u}_{\alpha,k}} V^{\textrm{int}}_{\beta,j}({\mathbf{u}})\right].
\end{equation}
The terms in lines 2-5 in Eq. \ref{EqMotion3} stem from the application of the derivative on the position dependent resolution function. The terms
$\mathbf{F}_{i}^\textrm{drift}$ are undesired forces that act only in the hybrid region, can lead to thermodynamic imbalances in the system, and, for example, artificially push particles from one subregion of the system to the other. In accordance with earlier works using the H-AdResS scheme, we will refer to these forces as \textit{drift forces} \cite{Potestio2013}. They need to be compensated, which can be achieved via the FEC, line 5 in Eq. \ref{EqMotion3}. In fact, the latter is typically constructed to cancel their average effect \cite{Potestio2013,Potestio2013a,Espanol2015}. 

The drift force $\mathbf{F}_{1}^\textrm{drift}$ comes from the potential energy interpolation and would not be present if we changed only the masses of the atoms but not the force field. $\mathbf{F}_{2}^\textrm{drift}$ is a result of choosing the kinetic masses of the higher modes to be adaptive and would be absent in the CKM approach. Finally, $\mathbf{F}_{3}^\textrm{drift}$ corresponds to the adaptive spring masses. As the resolution $\lambda_\alpha$ and the adaptive mass $\mu_\alpha$ depend only on the centroid positions of the ring polymers, drift forces only occur in the equations of motion for the centroid. Therefore, the internal motion of the ring polymers is not disturbed by any drift forces and can smoothly change from quantum to classical and vice versa when the atoms move through the system. Only the translation of the ring polymers is affected by drift forces, which can be corrected via the FEC. Note that the sums in $\mathbf{F}_{2}^\textrm{drift}$ and $\mathbf{F}_{3}^\textrm{drift}$ only run over non-centroid modes, since the centroid mode neither appears in the spring constant term in the Hamiltonian, nor is it associated with a variable mass in the kinetic energy onto which the position derivative could act.

\subsection{Integration} \label{methodology2_5_9}
To devise a suitable integration scheme for the equations of motion, Eqs. \ref{EqMotion1}-\ref{EqMotion4}, we use the Liouville operator formalism. To simplify the notation, in the following we will drop the atom index $\alpha$, as the Liouville operators for different ring polymers, commute and do not require further discussion. 

The Liouville operator $iL$ can be written as
\begin{equation}\label{iL1}
iL=iL^{(1)}+iL^{(2)},
\end{equation}
where $iL^{(1)}$ propagates positions and $iL^{(2)}$ momenta. We further decompose
\begin{eqnarray}\label{iL2}
iL^{(1)} & = iL^{(1)}_1 + iL^{(1)}_k,\\ 
iL^{(2)} & = iL^{(2)}_1 + iL^{(2)}_k,
\end{eqnarray}
where operators $iL^{(i)}_1$ propagate the first mode and $iL^{(i)}_k$ act only on the $k > 1$ modes. They are
\begin{eqnarray} \label{iL3_1}
iL^{(1)}_1 & = & P\frac{\mathbf{p}_1}{m}\frac{\partial}{\partial \mathbf{u}_1}, \\  \label{iL3_2}
iL^{(1)}_k & = & \frac{\mathbf{p}_k}{\tilde{\nu}(\mathbf{u}_1)}\frac{\partial}{\partial \mathbf{u}_k}, \\  \label{iL3_3}
iL^{(2)}_1 & = & \left( \mathbf{F}_1(\mathbf{u}) + \eta(\mathbf{u}_1) \sum_{k=2}^P \mathbf{p}_k^2 \right) \frac{\partial}{\partial \mathbf{p}_1},  \\   \label{iL3_4}
iL^{(2)}_k & = & \mathbf{F}_k(\mathbf{u}) \frac{\partial}{\partial \mathbf{p}_k},
\end{eqnarray}
with
\begin{equation}\label{iL4}
\begin{split}
\mathbf{F}_1(\mathbf{u}) & = \mathbf{F}^\textrm{H-AdResS}_{1}(\mathbf{u}) \\
& \qquad - \frac{1}{P}\left[ \sum_{i=1}^P  \left\{ V_i^{\textrm{QM}}(\mathbf{u}) - V_i^{\textrm{CL}}(\mathbf{u}) \right\} \right] \boldsymbol{\nabla}_{\mathbf{u}_1} \lambda(\mathbf{u}_1) \\
& \qquad - \left[ \sum_{i=2}^P \frac{1}{2} \xi_{i} (m-M) \omega_P^2 \mathbf{u}_{i}^2 \right] \boldsymbol{\nabla}_{\mathbf{u}_1} \lambda(\mathbf{u}_1) \\
& \qquad +\boldsymbol{\nabla}_{\boldsymbol{u}_{1}} \Delta H(\boldsymbol{u}_{1}),
\end{split}
\end{equation}
and
\begin{eqnarray} \label{iL5}
\mathbf{F}_k(\mathbf{u}) & = & -\nu_{k}(\mathbf{u}_1)\omega_P^2{\mathbf{u}}_{k}+\mathbf{F}^\textrm{H-AdResS}_{k}(\mathbf{u}), \\
\eta(\mathbf{u}_1) & =  &-  \frac{1}{2} \frac{(M-m)}{P\tilde{\nu}(\mathbf{u}_1)^2}  \boldsymbol{\nabla}_{\mathbf{u}_1} \lambda(\mathbf{u}_1).
\end{eqnarray}
The vector notation denotes that each Liouville operator in Eqs. \ref{iL3_1}-\ref{iL3_4} represents a set of three operators for each direction, which commute and can therefore be applied in arbitrary order.

In the first step we decompose the classical propagator $\exp\{iLt\}$ as
\begin{equation} \label{iL6}
e^{iLt}=e^{iL^{(1)}t+iL^{(2)}t}= \lim_{M\rightarrow\infty}\left( e^{iL^{(2)}\frac{t}{2M}} e^{iL^{(1)}\frac{t}{M}} e^{iL^{(2)}\frac{t}{2M}} \right)^M
\end{equation}
using the symmetric Trotter theorem \cite{Trotter1959,Strang1968}. Defining the time step $\Delta t=t/M$ this yields the velocity Verlet integrator
\begin{equation} \label{iL7}
e^{iL\Delta t}\approx  e^{iL^{(2)}\frac{\Delta t}{2}} e^{iL^{(1)}\Delta t} e^{iL^{(2)}\frac{\Delta t}{2}}.
\end{equation}
However, recalling the definitions of $iL^{(1)}$ and $iL^{(2)}$ we recognize that their constituents $iL^{(1)}_1$ and $iL^{(1)}_k$ as well as $iL^{(2)}_1$ and $iL^{(2)}_k$ do not commute. Therefore, using again the Trotter theorem, we decompose each of the propagators in Eq. \ref{iL7} further to
\begin{equation} \label{iL8}
\begin{split}
e^{iL\Delta t} & \approx  e^{iL^{(2)}\frac{\Delta t}{2}} e^{iL^{(1)}\Delta t} e^{iL^{(2)}\frac{\Delta t}{2}} \\
& \approx  \left( e^{iL^{(2)}_k\frac{\Delta t}{4}} e^{iL^{(2)}_1\frac{\Delta t}{2}} e^{iL^{(2)}_k\frac{\Delta t}{4}} \right) \left( e^{iL^{(1)}_1 \frac{\Delta t}{2}} e^{iL^{(1)}_k\Delta t} e^{iL^{(1)}_1\frac{\Delta t}{2}} \right) \left( e^{iL^{(2)}_k\frac{\Delta t}{4}} e^{iL^{(2)}_1\frac{\Delta t}{2}} e^{iL^{(2)}_k\frac{\Delta t}{4}} \right),
\end{split}
\end{equation}
which is correct up to second order in $\Delta t$ and is suited for the integration of the system's equations of motion. Note that for the CKM approach, the adaptive mass $\nu'_k(\mathbf{u}_1)$ in $iL^{(1)}_k$, Eq. \ref{iL3_2}, would be constant and also the $\mathbf{p}_k^2$ term in $iL^{(2)}_1$, Eq. \ref{iL3_3}, would be missing. Therefore, no second decomposition level would be required and we could stick with velocity Verlet.

The interpretation of the integration scheme in Eq. \ref{iL8} is straightforward. The first and the last term in brackets $(\cdots)$ correspond to the propagation of momenta, while the term in the middle propagates coordinates. The next decomposition level tells us how to update the momenta and the coordinates. The momenta are integrated in the following way: We first propagate all higher mode momenta by a quarter step, then we update the $\mathbf{p}_k^2$ term in $iL_1^{(2)}$ using these new momenta and propagate the first mode's momentum by a half step. Next, we perform the other quarter step for the higher modes. To get the full time step, the procedure is repeated after the position update in the center of the scheme. The coordinates are updated as follows: We first propagate the first mode by a half step using $iL_1^{(1)}$. Then we update the adaptive masses in the Liouville operator $iL_k^{(1)}$ and propagate the higher modes by a full step. Finally, we integrate the first mode by another half step. The scheme requires little additional computational overhead compared to a regular velocity Verlet scheme. The number of additional operations scales only linearly with the number of particles, and the force computation, usually the numerically most demanding part of a simulation, has to be performed as usual only once after all coordinates are fully propagated. 

Finally, we want to address the symplecticity of the new integrator. The Hamiltonian $H_P^{\textrm{akm}}$, Eq. \ref{EQ:17}, is not trivially separable into two parts, one depending only on coordinates and one containing only momenta. However, it has no term in which both the momentum and the corresponding conjugate coordinate of the same mode appear together. This is a result of our choice of the adaptive masses: The higher mode masses are position dependent, but they do not depend on their own mode coordinates but only on the centroid coordinate. However, the centroid itself, which determines the resolution and the masses of the ring polymers, is associated with a constant mass. Therefore, the Hamiltonian $H_P^{\textrm{akm}}$ and the corresponding equations of motion still define a symplectic structure. As a consequence, also the integration scheme in Eq. \ref{iL8} is symplectic, as it is constructed in a rigorous fashion from $H_P^{\textrm{akm}}$'s equations of motion using the Liouville operator formalism. This can also be understood considering the Liouville operators themselves: The operator $iL^{(1)}_k$, Eq. \ref{iL3_2}, which propagates the \textit{higher mode coordinates} $\mathbf{u}_k$ does have an additional position dependence but only on the \textit{centroid mode}. Hence, it can be applied as usual in a well-defined manner. Similarly, the operator $iL^{(2)}_1$, Eq. \ref{iL3_3}, propagating the momentum of the \textit{centroid mode} has an additional dependence on momenta, but only on the \textit{higher mode momenta}. Consequently, the determinant $J$ of the time evolution matrix is $1$. The symplecticity has the practical advantage that we are able to derive an energy conserving integrator, which, in our case, is exact up to second order in time, similar to a regular velocity Verlet.

It is worth pointing out that the previous observation is in contrast with earlier works using adaptive masses \cite{Rudnicki1994,Potestio2014a}. There, both momenta and the corresponding conjugate coordinates appear together in the same terms in the Hamiltonian. Hence, in those cases, Liouville's theorem no longer holds, the Liouville operator formalism breaks down, and symplecticity is lost.

\subsection{Multiple time-stepping} \label{methodology2_6_9}
In typical complex soft matter systems, non-bonded interactions as well as bonds, angles and dihedrals generate motion on different time scales. In PIMD, we have additionally the springs between the beads of the ring polymers onto which the quantum particles are mapped. If the kinetic masses for the higher modes are small, they vibrate strongly, which requires a small integration time step. When only sampling statistical averages, the kinetic masses can be chosen freely, for example, such that all higher modes vibrate with the same frequency. When calculating approximate quantum dynamical quantities, however, the kinetic mode masses must either correspond to the real ones, as in RPMD, or must be significantly decreased, as in CMD. This leads to an internal ring polymer dynamics which is significantly faster than the motion due to typical interatomic non-bonded or bonded potentials. Furthermore, in the CKM approach, the modes' oscillation frequencies are increased in the CL region, as the kinetic mass will be small there compared to the increased spring mass. This strongly motivates the introduction of multiple time-stepping into our integrator. 

We employ the RESPA scheme \cite{Tuckerman1992} and decompose the force computation into three parts: one for non-bonded forces, a second for the bonds, and a third for the internal ring polymer motion. The first drift term, $\mathbf{F}_1^\textrm{drift}$, depends only on the energies associated with the non-bonded potentials and is therefore evaluated together with the rest of the non-bonded forces. The second and the third drift terms $\mathbf{F}_2^\textrm{drift}$ and $\mathbf{F}_3^\textrm{drift}$, however, depend directly on the motion of the higher modes and therefore need to be evaluated together with them. Hence, we define
\begin{eqnarray} \label{iLtimestepping1_1}
iL^{(1)}_1 & = & P\frac{\mathbf{p}_1}{m}\frac{\partial}{\partial \mathbf{u}_1}, \\  \label{iLtimestepping1_2}
iL^{(1)}_k & = & \frac{\mathbf{p}_k}{\tilde{\nu}(\mathbf{u}_1)}\frac{\partial}{\partial \mathbf{u}_k}, \\  \label{iLtimestepping1_3}
iL^{(2)}_1 & = & \left( \mathbf{F}^{\textrm{mode}}_1(\mathbf{u}) + \eta(\mathbf{u}_1)\sum_{k=2}^P  \mathbf{p}_k^2 \right) \frac{\partial}{\partial \mathbf{p}_1},  \\   \label{iLtimestepping1_4}
iL^{(2)}_k & = & \mathbf{F}^{\textrm{mode}}_k(\mathbf{u}) \frac{\partial}{\partial \mathbf{p}_k}, \\ \label{iLtimestepping1_5}
iL^{(3)}_q & = & \mathbf{F}_q^{\textrm{int}}(\mathbf{u}) \frac{\partial}{\partial \mathbf{p}_q}, \\
\label{iLtimestepping1_6}
iL^{(4)}_q & = & \mathbf{F}_q^{\textrm{nb}}(\mathbf{u}) \frac{\partial}{\partial \mathbf{p}_q},
\end{eqnarray}
with
\begin{eqnarray}\label{iLtimestepping2_1}
\mathbf{F}^{\textrm{mode}}_1(\mathbf{u}) & = & - \left[ \sum_{i=2}^P \frac{1}{2} \xi_{i} (m-M) \omega_P^2 \mathbf{u}_{i}^2 \right] \boldsymbol{\nabla}_{\mathbf{u}_1} \lambda(\mathbf{u}_1), \\ \label{iLtimestepping2_2}
\mathbf{F}^{\textrm{mode}}_k(\mathbf{u}) & = & -\nu_{k}(\mathbf{u}_1)\omega_P^2{\mathbf{u}}_{k}, \\ \label{iLtimestepping2_3}
\mathbf{F}^{\textrm{int}}_q(\mathbf{u}) & = & -\frac{1}{P}\sum_{j = 1}^P\ \sum_{\beta = 1}^N \boldsymbol{\nabla}_{\boldsymbol{u}_q} V^{\textrm{int}}_{\beta,j}({\mathbf{u}}), \\ \label{iLtimestepping2_4}
\mathbf{F}^{\textrm{nb}}_q(\mathbf{u}) & = & -\frac{1}{P}\sum_{j = 1}^P\ \sum_{\beta = 1}^N \left[ \lambda_{\beta} \boldsymbol{\nabla}_{\boldsymbol{u}_q} V^{\textrm{QM}}_{\beta,j}({\mathbf{u}}) + \left(1 - \lambda_{\beta}\right) \boldsymbol{\nabla}_{\boldsymbol{u}_q} V^{\textrm{CL}}_{\beta,j}({\mathbf{u}})\right] \\ \nonumber
&& \qquad - \frac{1}{P}\left[ \sum_{i=1}^P  \left\{ V_i^{\textrm{QM}}(\mathbf{u}) - V_i^{\textrm{CL}}(\mathbf{u}) \right\} \right] \boldsymbol{\nabla}_{\mathbf{u}_q} \lambda(\mathbf{u}_1) \\ \label{iLtimestepping2_5}
&& \qquad +\boldsymbol{\nabla}_{\boldsymbol{u}_{q}} \Delta H(\boldsymbol{u}_{1}). \nonumber
\end{eqnarray}
Note that the Liouville operators $iL_q^{(3)}$ and $iL_q^{(4)}$ do not need to be split into centroid and higher terms, as these commute in this case. Hence, for the sake of brevity, we have subsumed both parts and changed to the index $q$, which includes all $1 \leq q \leq P$. Finally, we obtain the following RESPA multiple time-stepping scheme:
\begin{equation} \label{iLtimestepping3}
\begin{split}
e^{iL\Delta t} & \approx  e^{iL^{(4)}\frac{\Delta t}{2}} \Biggl\{ e^{iL^{(3)}\frac{\delta t}{2}} \Biggl[ \left( e^{iL^{(2)}_k\frac{dt}{4}} e^{iL^{(2)}_1\frac{dt}{2}} e^{iL^{(2)}_k\frac{dt}{4}} \right) \left( e^{iL^{(1)}_1 \frac{dt}{2}} e^{iL^{(1)}_k dt} e^{iL^{(1)}_1\frac{dt}{2}} \right) \times \\
& \qquad\quad \times\left( e^{iL^{(2)}_k\frac{dt}{4}} e^{iL^{(2)}_1\frac{dt}{2}} e^{iL^{(2)}_k\frac{dt}{4}} \right) \Biggr]^n e^{iL^{(3)}\frac{\delta t}{2}} \Biggr\}^N e^{iL^{(4)}\frac{\Delta t}{2}},
\end{split}
\end{equation}
where $\Delta t = N\cdot\delta t=N\cdot n\cdot dt$. The internal ring vibrations as well as the drift terms depending on these higher ring modes are integrated with the shortest time step $dt$. The intramolecular bonds and angles are integrated with a distinct, intermediate time step $\delta t$, and the intermolecular non-bonded interactions as well as the corresponding drift terms are integrated with the largest time step $\Delta t$. The whole integration may be carried out in normal mode space, although in practice the interatomic forces are computed in real space and then transformed into mode space.

\subsection{Langevin thermostating} \label{methodology2_7_9}
To generate a canonical ensemble we need to couple the system to a thermostat. We resort to a Langevin thermostat, as Langevin equation-based frameworks have been shown to be favorable in PIMD and RPMD simulations and can be used to optimize sampling efficiency \cite{Ceriotti2009,Ceriotti2010,Ceriotti2011,Ceriotti2012}. As the focus of this work is not advanced thermostating, however, we use a simple white noise Langevin thermostat without memory instead of, for example, a GLE approach. The implementation follows the BAOAB method by Leimkuhler and Matthews \cite{Leimkuhler2013}, which provides high configurational sampling accuracy. Within the proposed multiple time-stepping scheme this yields
\begin{equation}\label{iLtimestepping4}
\begin{split}
e^{iL\Delta t} & \approx  e^{iL^{(4)}\frac{\Delta t}{2}} \Biggl\{ e^{iL^{(3)}\frac{\delta t}{2}} \Biggl[ \left( e^{iL^{(2)}_k\frac{dt}{4}} e^{iL^{(2)}_1\frac{dt}{2}} e^{iL^{(2)}_k\frac{dt}{4}} \right) \left( e^{iL^{(1)}_1 \frac{dt}{4}} e^{iL^{(1)}_k \frac{dt}{2}} e^{iL^{(1)}_1\frac{dt}{4}} \right) \times \\
& \qquad \times e^{iL^{\textrm{Langevin}} dt} \left( e^{iL^{(1)}_1 \frac{dt}{4}} e^{iL^{(1)}_k \frac{dt}{2}} e^{iL^{(1)}_1\frac{dt}{4}} \right) \left( e^{iL^{(2)}_k\frac{dt}{4}} e^{iL^{(2)}_1\frac{dt}{2}} e^{iL^{(2)}_k\frac{dt}{4}} \right) \Biggr]^n e^{iL^{(3)}\frac{\delta t}{2}} \Biggr\}^N e^{iL^{(4)}\frac{\Delta t}{2}},
\end{split}
\end{equation}
with the action of the Langevin Liouville operator $iL^{\textrm{Langevin}}$ on mode $i$
\begin{eqnarray} \label{iLtimestepping5_1}
e^{iL^{\textrm{Langevin}} dt}\,\mathbf{u}_i & = & \mathbf{u}_i, \\ \label{iLtimestepping5_2}
e^{iL^{\textrm{Langevin}} dt}\,\mathbf{p}_i & = & \mathbf{p}_i\,e^{-\gamma dt}+\sqrt{\nu'_i(\mathbf{u}_1) k_bT (1-e^{-2\gamma dt})}\,\mathbf{R}(t).
\end{eqnarray}
$\gamma$ is the friction parameter, $T$ the temperature, $k_B$ Boltzmann's constant, and $R(t)$ are independent and identically distributed normal random numbers with mean 0, variance 1, and $\langle \mathbf{R}(t)\mathbf{R}(t')\rangle=\boldsymbol{\delta}(t-t')$. This thermostating method can also be adapted such that each mode is thermostated with a different optimized friction constant, as done in the path integral Langevin equation (PILE) scheme by Ceriotti et al. \cite{Ceriotti2010,Rossi2014}. 

Using the integration scheme of Eq. \ref{iLtimestepping4}, we can perform efficient adaptive quantum--classical PIMD simulations with either the AKM or the CKM approach. It is derived in a rigorous fashion from a symplectic Hamiltonian and is also consistent with PI quantization, provided that the criteria in Eqs. \ref{EQ:6} and \ref{EQ:7} are satisfied. It is computationally advantageous over full-quantum simulations, because in the CL region all forces between interacting ring polymers can be approximated by a single calculation between the centroids. Furthermore, because the ring polymers are collapsed in the CL region and interact classically, the integration of the internal motion, i.e. of the higher modes, can be stopped and the ring polymers can be frozen in this part of the system. In the CKM case or for full-quantum systems, the algorithm reduces to a regular velocity Verlet scheme with multiple time-stepping and Langevin thermostating.

The derived quantum--classical multiresolution scheme can be combined with other optimization techniques for PI simulations. For example, the non-bonded forces could also in the QM region be evaluated based on fewer than $P$ beads, using the RPC scheme by Markland and Manolopoulos \cite{Markland2008,Markland2008a}. Alternatively, instead of white noise Langevin thermostating, we could make use of a colored noise thermostat. It was shown by Ceriotti et al. that a carefully parametrized PI GLE can lower the number of Trotter beads required for converged quantum behavior \cite{Ceriotti2009,Ceriotti2010,Ceriotti2011,Ceriotti2012}. As our approach reduces the overall computational effort of a PI simulation by restricting the QM region of the system, it is complementary to these methods, which reduce the numerical complexity of the PI interactions themselves.

\section{Approximate quantum dynamics} \label{methodology3_9}
In PIMD, we only measure quantum statistical properties. In the following, we will discuss how our integration scheme can be extended to allow for multiresolution quantum--classical CMD and RPMD with only minor changes.

\subsection{Quantum--classical centroid molecular dynamics} \label{methodology3_1_9}
Centroid molecular dynamics (CMD) is a method for the calculation of real-time quantum correlation functions in the short-time limit \cite{Cao1994d,Cao1994c,Cao1994e,Cao1994,Cao1994b,Cao1996,Martyna1996,Cao1996a,Jang1999,Hone2006,Perez2009,Polyakov2010}. It is based on the notion that approximate quantum dynamical properties can be calculated from the time evolution of the centroid subject to the potential of mean force generated by the ring polymer. Formally, this potential is obtained by integration over all possible ring configurations with constrained centroid position. This would be not just computationally expensive but practically intractable. The idea of CMD is to adiabatically decouple the internal fluctuations of the ring polymers from the centroid motion. By rescaling the higher mode kinetic masses ($k>1$) with a sufficiently small adiabadicity parameter $0 < \gamma_\textrm{CMD}^2 < 1$, such that $\nu'_k\rightarrow\gamma_\textrm{CMD}^2\nu'_k$, the higher modes can be forced to evolve significantly faster than the centroid. Thereby, the centroid potential of mean force of the ring polymer is generated ``on the fly'' during the simulation. It has been shown, however, that in practice a partial adiabatic decoupling is sufficient for most applications \cite{Hone2006}. In addition to the mass rescaling, the higher modes alone are coupled to thermostats such that the centroid dynamics remains Newtonian.

The previously described implementation of CMD, i.e. the removal of the thermostat from the centroid and the kinetic mass rescaling, can also be done easily in our quantum--classical multiresolution scheme. In practice, one would typically be interested in the quantum dynamics only in the QM region. Hence, it suffices to only remove the centroid thermostat in this region. Then we could measure approximate quantum dynamical properties in the QM region while the classical part would still behave as in the canonical ensemble and could serve, for example, as a particle reservoir for the QM region. Another relevant scenario is the simulation of a complex biomolecule like a protein. In this case, an overall large simulation box would be required to preserve the structure and the solvating environment of the system, although we may want to probe the dynamics only in a specific subregion, such as near the protein's active site~\cite{Wang2014}.

\subsection{Quantum--classical ring polymer molecular dynamics} \label{methodology3_2_9}
An alternative approach to the calculation of approximate quantum dynamics is provided by ring polymer molecular dynamics (RPMD) \cite{Craig2004,Braams2006,Hone2006,Perez2009,Habershon2013}. In the normal RPMD approach, the kinetic masses are chosen to be the real physical masses (as we did above also for PIMD) and no thermostats are used, such that the ring polymer evolution is completely Newtonian. In comparison to CMD, RPMD uses the whole chain to approximate quantum time correlation functions. However, the internal ring fluctuations can lead to artifacts when measuring, for example, vibrational spectra \cite{Witt2009}. To overcome this deficiency, Rossi, et al. recently proposed a thermostated ring polymer molecular dynamics (TRPMD) approach \cite{Rossi2014} that can be interpreted as an intermediate method between normal RPMD and CMD. In TRPMD, the kinetic masses are also chosen to be the real physical masses and measurements are performed based on the whole chain. However, as in CMD, Langevin thermostats are attached to all higher modes for which $k>1$. Provided the thermostats are adjusted carefully,
TRPMD avoids both the spurious resonances in the vibrational spectra and also the curvature problem of CMD \cite{Witt2009}, while retaining the appealing properties of RPMD. An ideal choice for the Langevin friction parameters in TRPMD is given by the PILE scheme \cite{Ceriotti2010,Rossi2014}. In the PILE method, each higher mode $k>1$ is thermostated with a different optimized coupling constant $\gamma_k$ based on the mode vibration frequency as $\gamma_k=\omega_P\sqrt{\xi_kP}$.

Just as with CMD, TRPMD simulations can be easily run with our quantum--classical PI scheme and the corresponding integrator, Eq. \ref{iLtimestepping4}. We only need to adapt the thermostats on the different modes accordingly and remove the thermostat on the centroid. 

Note that in the AKM approach the kinetic masses will change in the CL region. However, in this part of the simulation box the spring masses also have different values and we are typically not interested in the dynamics anyway. Thus, one may also reintroduce the centroid thermostat in this outer region.

\section{Simulations} \label{simulations_9}
To validate the proposed adaptive resolution PIMD approach, we implement it in the ESPResSo++ molecular simulation package \cite{Halverson2013} and perform adaptive resolution simulations of liquid water. Nuclear quantum effects in liquid water have been thoroughly investigated and shown to be important for an accurate description of its structure and dynamics \cite{Morrone2008,Paesani2010,Ceriotti2013,Fritsch2014}. Hence, water is an ideal test case for the method.

\subsection{Water system} \label{simulations_9_1}
We consider a system of 918 water molecules in a slab-shaped box with dimensions $L_x = 6.92\,\text{nm}$, $L_y=L_z=2.0\,\text{nm}$ ($33.168\,$molecules/nm${}^3$), and periodic boundary conditions in all directions. The resolution changes along the $X$-direction, the full width of the QM region is set to $d_{\text{QM}}=2.0\,\text{nm}$, and the width of the adjacent HY regions to $d_{\text{HY}}=1.5\,\text{nm}$. The resolution function is given by a squared cosine, commonly used in adaptive resolution simulations \cite{Potestio2012,Espanol2015,Fogarty2015,Kreis2016,Kreis2016b,Kreis2016c,Peters2016}. We perform the simulations at a temperature of $300\,K$ and use a Trotter number $P=32$, as this has been shown to provide well-converged results for most dynamical and structural water properties \cite{Morrone2008,Paesani2010,Fritsch2014}. A simulation snapshot of the system is presented in Fig. \ref{fig1_9} (a). 

To model the water, we use a force field that was recently developed by Fritsch et al. specifically for PI simulations of bulk liquid water \cite{Fritsch2014}. It is parametrized from \textit{ab initio} density functional theory calculations using the force matching \cite{Ercolessi1994,Izvekov2005a,Izvekov2005} and iterative Boltzmann inversion methods \cite{Reith2003}. All interactions are mapped onto a set of short-ranged tabulated potentials and no explicit charges are present. Separate potentials are provided for the non-bonded O-O, O-H, H-H interactions, for the O-H bond, and for the H-O-H angle. An additional bonded potential is applied between the two H-atoms of the same molecule. This force field describes the structural and dynamical properties of liquid water at $300\,K$ and at a density of $1.1$ g/cm${}^2$ very well. Furthermore, it is very efficient in simulations, since it is purely short-ranged with an interaction cutoff of $7.8\,$\AA. We have chosen this potential for its numerical efficiency and its suitability for PI simulations, and we note that the derived adaptive resolution methodology can also be applied for analytic potentials as well as those that include charges.

In order to collapse the ring polymers in the CL region, we choose $M_\alpha=100\,m_\alpha$ for all particles $\alpha$. Because of their point-like structure we only use the centroids to calculate non-bonded and bonded interactions between atoms in the CL region (see Fig. \ref{fig1_9} (b)). Furthermore, we stop the integration of the higher modes in the CL region, i.e., we freeze the internal degrees of freedom of the ring polymers. Note that the setup satisfies the criteria in Eqs. \ref{EQ:6} and \ref{EQ:7} and can therefore be considered to be consistent with formal path integral quantization.

We run simulations using both the AKM and CKM approaches, although we focus on the AKM method, for which we have derived a non-standard integrator and which allows larger time steps. In all simulations, the kinetic mass of the centroids is given by Eq. \ref{kineticmasses1}, which corresponds to using the real mass. For the AKM simulations, we choose the kinetic masses of the higher modes according to Eq. \ref{kineticmasses2}. As already argued, this corresponds to using the real masses in the QM region, which facilitates a realistic ring polymer time evolution and therefore allows the calculation of approximate quantum dynamical properties from RPMD simulations. In the CL region, the higher mode masses are increased as explained previously. For the CKM simulations, we choose the masses in a similar way, although they remain constant over all simulation domains. For CMD simulations, we introduce an additional rescaling of the higher modes' kinetic masses with an adiabadicity parameter $\gamma^2_\textrm{CMD}=0.05$. Note that we do not rescale the kinetic masses of the higher modes with the eigenvalues of the normal mode transformation, as it is often done in CMD \cite{Hone2006}. We also keep the additional $1/P$ factor in the kinetic masses, which we introduced earlier to ensure dynamics on the correct time scale. Therefore, the ring polymers' higher modes vibrate with frequencies $\omega_P\sqrt{\xi_kP/\gamma^2_\textrm{CMD}}$.

To enforce the correct temperature, we couple all modes to white noise Langevin thermostats. The centroid mode is thermostated with a friction constant $\gamma=2.0\,\textrm{ps}^{-1}$, except in CMD and TRPMD simulations, where no thermostat is applied on it. For the higher modes $k$, we employed the PILE scheme by Ceriotti et al. \cite{Ceriotti2010,Rossi2014} and used frictions $\gamma_k=\omega_P\sqrt{\xi_kP}$ that are proportional to the modes' vibration frequencies (for CMD simulations, we used $\gamma_k=\omega_P\sqrt{\xi_kP/\gamma^2_\textrm{CMD}}$). The PILE method leads to optimized sampling and can also be applied in the context of TRPMD simulations.

The derived adaptive quantum--classical simulation method allows to not only change the quantum delocalization of the particles but also their non-bonded interaction potentials. This has been demonstrated in our previous paper in simulations of liquid parahydrogen \cite{Kreis2016}. Here, we perform the majority of the validation simulations using the same interaction potential in both the QM and the CL region. We also test the scheme using a different potential in the CL region, a purely repulsive Weeks-Chandler-Andersen (WCA) potential \cite{JohnD.WeeksDavidChandler1971} of the form
\begin{equation}
V^\textrm{CL}(r) =
\left\{
	\begin{array}{ll}
		4\epsilon\Bigl[\left(\frac{\sigma}{r}\right)^{12}-\left(\frac{\sigma}{r}\right)^6+\frac{1}{4}\Bigr]  & \mbox{: } r \leq R_{\text{c}} \\
		0 & \mbox{: } r > R_{\text{c}}
	\end{array}
\right.
\end{equation}
with $\epsilon=k_BT$, $\sigma =0.25\,\text{nm}$, and $R_{\text{c}}=2^{\frac{1}{6}}\sigma=0.28\,\text{nm}$. The potential acts only between the oxygen atoms. Note that the intramolecular bonded interactions are kept in the CL region to prevent the molecules from disintegrating. 

\subsection{Setups} \label{simulations_9_2}
We perform simulations employing the following setups:
\begin{itemize}
\item \textbf{Setup 1:} We use adaptive kinetic masses and the same interaction potentials in both regions. Applying thermostats to all modes, we calculate various structural properties of the water in the QM region. Additionally, we remove the thermostat from the centroid and use TRPMD to calculate several dynamical quantities.
\item \textbf{Setup 2:} The same as setup 1, except the kinetic masses of the higher modes are rescaled with the adiabadicity parameter $\gamma^2_\textrm{CMD}$. Then, we calculate the dynamical properties via CMD.
\item \textbf{Setup 3:} The AKM method is applied as in setup 1, but the WCA potential is employed to model intermolecular interactions in the CL region. In this scenario, we only validate the coupling by calculating density profiles as well as profiles of the atomistic radii of gyration.
\item \textbf{Setup 4:} We switch to constant kinetic masses and employ the same interaction potentials in both regions. As in setup 3, we only validate the coupling and calculate density profiles as well as profiles of the atomistic radii of gyration.
\end{itemize}

Additionally, we perform full-quantum and full-classical ($P=1$) reference simulations without any interpolation between different particle masses or interaction potentials. All simulation parameters, including the box dimensions, are the same as for the adaptive simulations. The only exception is the friction constant of the Langevin thermostat, which is set to $10.0\,\textrm{ps}^{-1}$ in the full-classical simulations. 

The time steps used in the simulation setups are presented in Tab. \ref{table_timesteps}. 
\begin{table}[!ht]\renewcommand{\arraystretch}{1.3}\addtolength{\tabcolsep}{-1pt}\renewcommand{\tabcolsep}{0.3cm}
\begin{center}
\begin{tabular}{c c c c}
\hline
Setup & $\Delta t$ & $\delta t$ & $dt$  \\ 
\hline
\#1: AKM method, same potentials, TRPMD & $2.0\,\text{fs}$ & $0.5\,\text{fs}$ & $0.05\,\text{fs}$ \\
\#2: AKM method, same potentials, CMD & $0.4\,\text{fs}$ & $0.1\,\text{fs}$ & $0.01\,\text{fs}$  \\
\#3: AKM method, WCA potential in CL region, TRPMD & $2.0\,\text{fs}$ & $0.5\,\text{fs}$ & $0.05\,\text{fs}$  \\
\#4: CKM method, same potentials, TRPMD & $1.0\,\text{fs}$ & $0.1\,\text{fs}$ & $0.00625\,\text{fs}$ \\
\hline
\end{tabular}
\caption[Time steps in the quantum--classical adaptive resolution path integral molecular dynamics simulations of liquid water.]{Time steps for the quantum--classical adaptive resolution PIMD, RPMD and CMD simulations and for the reference calculations.}
\label{table_timesteps}
\end{center}
\end{table}
For the full-classical and the full-quantum reference simulations we use the same time steps as in the corresponding adaptive resolution setups. The time steps in the table refer to those used in equilibration simulations, during the derivation of the free energy correction and the thermodynamic force (see next section), as well as during all other simulations sampling statistical averages. For the calculation of dynamic quantities in setup 1 we reduce all time steps to the same ones as used in the CMD simulations in setup 2. We do this for two reasons: On the one hand, our implementation of the integration scheme allows one to print out positions or velocities only after a full step $\Delta t$. Hence, this large time step needs to be short enough to allow a fine sampling when calculating, for example, velocity autocorrelation functions. On the other hand, we want to avoid artifacts resulting from the use of different time steps when comparing CMD to TRPMD. Note, however, that only few and very short simulations need to be run with this modification. The majority of simulations use the time steps in Tab. \ref{table_timesteps}.

In general, all time steps are chosen to be as large as possible but still sufficiently small to accurately sample phase space, retain an acceptable level of energy-conservation in microcanonical test simulations, and generate the correct temperature in simulations in the canonical ensemble. The time steps we find to work well seem reasonable: In classical simulations, updating the regular non-bonded forces every $1$-$2\,\textrm{fs}$ is a frequent choice \cite{Berendsen1987,Wu2006,Fogarty2014,Salvalaglio2015,Kreis2016b,Kreis2016c}, while the vibration frequency for the bonds and angles in water demands a time step of around $0.5\,\textrm{fs}$ \cite{Toukan1985,Fritsch2014}. The vibration frequency of the springs between the PI beads is yet higher, requiring an even smaller time step. Furthermore, CMD simulations are known to require particularly small time steps, as the internal ring polymer motion is strongly accelerated. A similar effect is observed in simulations with the CKM approach. In this case, the internal motion of collapsed ring polymers is also significantly enhanced (we mentioned that in the CL region, the ring polymers are frozen. However, the ring polymers are already strongly collapsed at the outer parts of the HY region, where a full integration of the internal motion is still necessary to accommodate the gradual collapse and extension of the ring polymers). Therefore, it becomes clear that the AKM scheme is better suited for the proposed adaptive quantum--classical simulation protocol than the naive CKM method. We want to stress, however, that finding optimal time steps is not the primary goal of this work and that there is certainly room for further fine-tuning.

For the calculation of all structural quantities and statistical averages we run simulations of duration $200\,\text{ps}$, if not otherwise indicated. Additionally, we perform short $2\,\text{ps}$ runs during which we calculate velocity autocorrelation functions and vibrational spectra. We also measure hydrogen bond population fluctuations, which is done in simulations of duration $32\,\text{ps}$. In all cases we start from equilibrated configurations, run 10 independent simulations, and average the results.

\subsection{Free energy corrections} \label{simulations_9_3}
To correct for the thermodynamic imbalance between the low-mass QM and the high-mass CL region, we apply a free energy correction (FEC) $\Delta H$ \cite{Potestio2013,Potestio2013a,Kreis2014,Espanol2015,Kreis2016,Kreis2016b}. We derive the FEC via Kirkwood thermodynamic integration (KTI) \cite{Kirkwood1935a} between the fully CL ($\lambda=0$) and the fully QM ($\lambda=1$) system and we calculate the averages
\begin{eqnarray} \label{KTI_term1}
1.&& \left\langle\frac{1}{NP} \sum_{\alpha=1}^N \sum_{i=1}^P  \left\{ V^{\textrm{QM}}_{\alpha,i}(\boldsymbol{u}) - V^{\textrm{CL}}_{\alpha,i}(\boldsymbol{u}) \right\} \right\rangle_\lambda, \\ \label{KTI_term2}
2.&& \left\langle\frac{1}{N} \sum_{\alpha=1}^N \sum_{i=2}^P \frac{1}{2} \frac{(M_\alpha-m_\alpha)}{P\tilde{\nu}_{\alpha}(\boldsymbol{u}_{\alpha,1})^2} \boldsymbol{p}_{\alpha,i}^2 \right\rangle_\lambda, \\ \label{KTI_term3}
3.&& \left\langle\frac{1}{N} \sum_{\alpha=1}^N \sum_{i=2}^P \frac{1}{2} \xi_{i} (m_\alpha-M_\alpha) \omega_P^2 {\mathbf{u}}_{\alpha,i}^2 \right\rangle_\lambda,
\end{eqnarray}
as well as the pressure $p(\lambda)$ for a set of 101 $\lambda$'s along the integration path from $\lambda=0$ to $\lambda=1$. 
The KTI is run in a smaller box of dimensions $L_x = L_y = L_z = 3.0\,\textnormal{nm}$. All other simulation parameters are as explained above, except for the thermostat friction of the centroid mode, which was set to $10\,\textnormal{ps}^{-1}$ to achieve rapid equilibration after changing $\lambda$. We start the KTI from an equilibrated system at $\lambda=0$ and we perform for each $\lambda$ a short $0.3\,\text{ps}$ equilibration run (for setup 2 only $0.12\,\text{ps}$ due to the short time step) and another $1.5\,\text{ps}$ run (for setup 2 only $0.6\,\text{ps}$) during which we take measurements. From these results we construct the FEC $\Delta H$ to cancel the averages of the drift forces, Eq. \ref{EqMotion3}, and the pressure difference between the subsystems. Since calculating the FEC via KTI is an approximate method to correct for the thermodynamic imbalance, we refine the FEC using the thermodynamic force (TF) scheme \cite{Fritsch2012,Kreis2016}. The TF is an iterative approach that directly constructs a correction force in the HY region from the distorted density profile along the direction of resolution change in order to flatten the density throughout the system. 
Each TF iteration consists of a $50\,\text{ps}$ equilibration run ($10\,\text{ps}$ for setup 2 and $15\,\text{ps}$ for setup 4) and a $150\,\text{ps}$ production run ($30\,\text{ps}$ for setup 2 and $20\,\text{ps}$ for setup 4) during which we sample the density. We perform 20 iterations for each setup. 

Although we calculate the quantities in Eqs. \ref{KTI_term1}-\ref{KTI_term3} separately for the oxygen and hydrogen atoms, we determine global molecular pressures instead of species-wise partial pressures. After constructing the correction force from the pressure $p(\lambda)$, we distribute it between oxygen and hydrogens proportionally to their masses. Finally, the FEC is applied on the atomistic level based on the atom's centroid positions.

Both the derivation of the FEC via KTI and the iterative correction are well-established methods for achieving a smooth coupling in adaptive resolution simulations. See, for example, Refs. \cite{Fritsch2012,Potestio2013,Potestio2013a,Kreis2014,Espanol2015,Kreis2016,Kreis2016b} for further technical details.

\section{Results} \label{results_9}
\subsection{Structure}
\begin{figure}[ht!]
\centering
  \includegraphics[width=0.75\columnwidth]{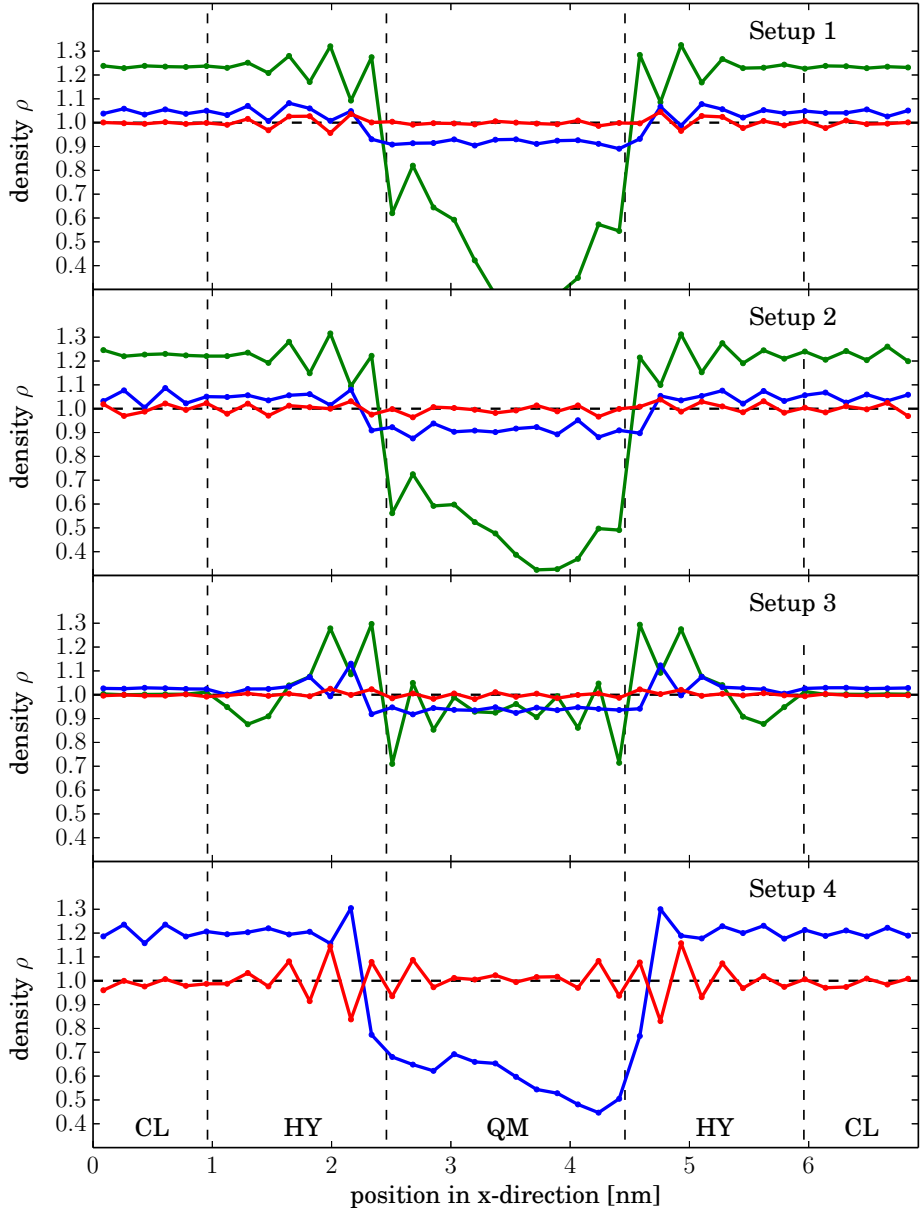}
  \caption[Density profiles in the quantum--classical adaptive resolution simulations of liquid water.]{Normalized density profiles $\rho$ in the quantum--classical adaptive resolution simulations of liquid water for the four different setups without FEC (green), with KTI-based FEC but without iterative refinement (blue), and with FEC including iterative refinement via TF (red). For setup 4, we are not able to run stable simulations without any FEC (setup 1: AKM, same potentials, TRPMD. Setup 2: AKM, same potentials, CMD. Setup 3: AKM, WCA potential in CL region, TRPMD. Setup 4: CKM, same potentials, TRPMD).
\label{chapter7_fig2}}
\end{figure}

We first investigate the structural properties of the adaptive quantum--classical water systems. Fig. \ref{chapter7_fig2} shows the density profiles along the x-direction of the four setups without correcting for the thermodynamic imbalance (green curves), with FEC but without iterative refinement (blue curves), and with FEC including the iterative refinement via TF (red curves).
Without any corrections the density is strongly distorted. Applying the non-iterative FEC significantly improves the coupling between the regions, although the density in the QM region is still slightly too low for setups 1-3 and much too low for setup 4. This can be expected, as the non-iterative FEC is an approximate method and since statistical inaccuracies can occur during its derivation via KTI. Refining the FEC with the iterative TF technique, we obtain flat density profiles for all setups, except setup 4, for which significant deviations in the HY region remain. Note that for setup 4, which uses the CKM scheme, we were not able to run stable simulations without any compensation. In this case, the drift forces are so strong that all molecules are immediately pushed to one subregion. In comparison, the AKM approach works much better and requires a more moderate FEC.

The derivation of the FEC via KTI and several iterations of TF may seem cumbersome. However, both the KTI as well as the TF iterations can be run using simulation setups that are much smaller than the actual system. For large applications this step will likely take significantly less time than the simulation of the complete system. Additionally, more advanced approaches have recently been developed that efficiently calculate the FEC on the fly during the simulation of the full system or a representative smaller one \cite{Heidari2016}. Note that a FEC or a similar compensation force is required in all adaptive resolution methods that allow a free exchange of particles between subregions that feature different thermodynamics \cite{Fritsch2012,Potestio2013,Potestio2013a,Agarwal2015,Agarwal2016,Heidari2016}. All results reported below are calculated in setups in which the refined FEC is applied.

\begin{figure}[ht!]
\centering
  \includegraphics[width=0.75\columnwidth]{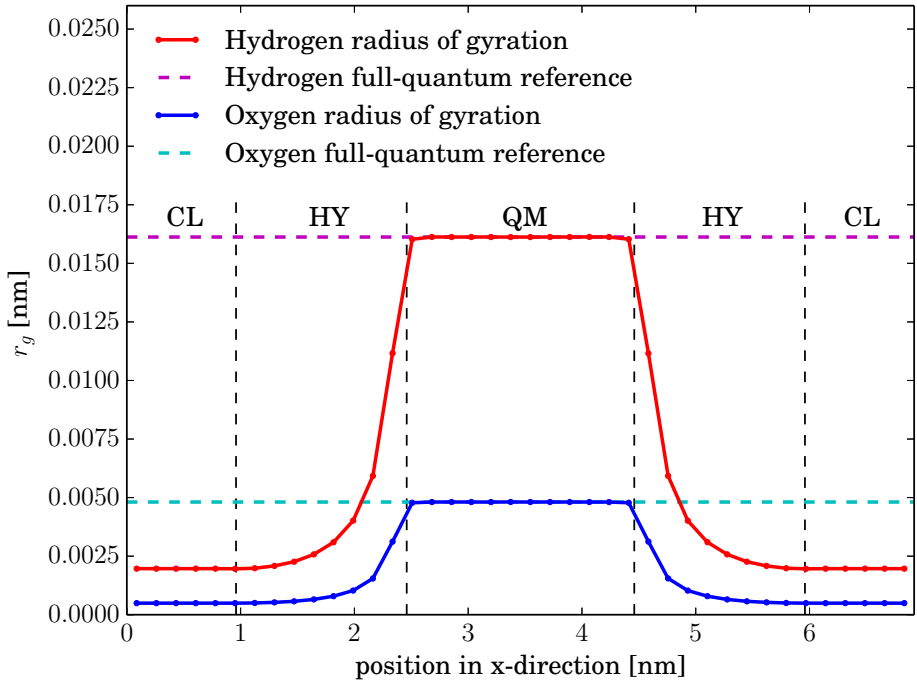}
  \caption[Radius of gyration profiles in the quantum--classical adaptive resolution simulations of liquid water.]{Radii of gyration of the ring polymers corresponding to oxygen and hydrogen atoms as a function of the ring polymers' position along the x-direction. The magenta and cyan lines correspond to the radii of gyration of atoms in full-quantum reference simulations. \label{chapter7_fig3}}
\end{figure}

Fig. \ref{chapter7_fig3} presents the radii of gyration of the ring polymers corresponding to the water's oxygen and hydrogen atoms as a function of their position along the x-direction. 
In the QM region, the radii of gyration perfectly match with those from full-quantum reference simulations, while in the CL region they drop by $\approx 90\%$ (also see Fig. \ref{fig1_9}). Therefore, the molecules exhibit their full-blown ``quantumness'' in the QM region, while in the CL region the ring polymers shrink to nearly point-like particles and behave classically. To collapse the ring polymers even further, one would simply need to choose a heavier particle mass $M_\alpha$ in the CL region. The radius of gyration in the CL region is approximately proportional to $1/\sqrt{M_\alpha}$. Note that the data in Fig. \ref{chapter7_fig3} correspond to system 1 and that the other setups show the exact same behavior.

\begin{figure}[ht!]
\centering
  \includegraphics[width=0.6\columnwidth]{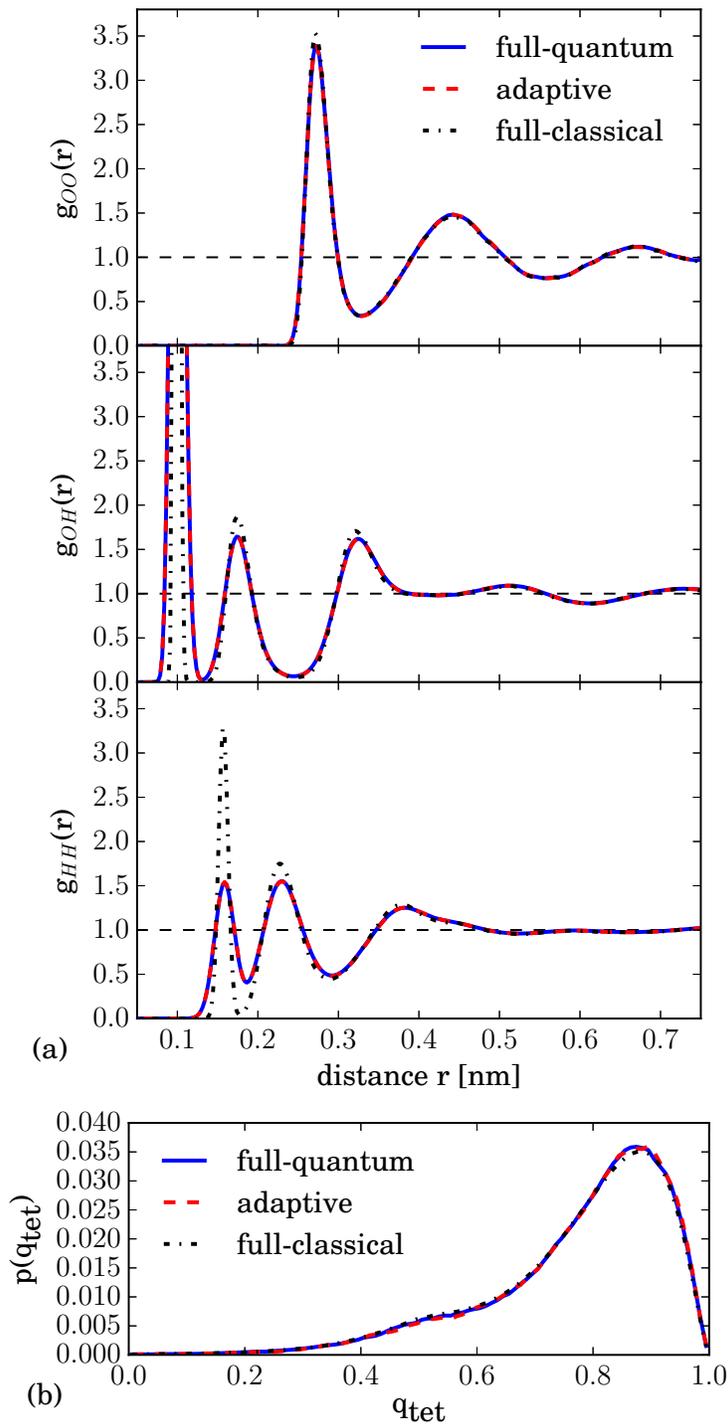}
  \caption[Radial distribution functions and tetrahedral order parameter in the QM region of quantum--classical adaptive resolution simulations of liquid water.]{(a) (Bead-bead) RDFs of the quantum--classical adaptive resolution simulations calculated in the QM region, and of full-quantum and full-classical ($P=1$) reference simulations. (b) Same for the tetrahedral order parameter $q_\textrm{tet}$. \label{chapter7_fig4}}
\end{figure}

Using setup 1, we also calculate the water's radial distribution functions (RDFs) and the tetrahedral order parameter $q_\textrm{tet}$ within the QM region (Fig. \ref{chapter7_fig4}).
We left a small buffer of $0.25\,\textrm{nm}$ at the interface to the HY region and considered the inner $1.5\,\textrm{nm}$ of the QM region in order to avoid artifacts by molecules at the outer edges of the QM region that interact strongly with molecules in the HY region. For a molecule $i$, $q_\textrm{tet}$ is given by
\begin{equation}
q_\textrm{tet} = 1- \frac{3}{8}\sum_{j=1}^3\sum_{k=j+1}^4\left(\textrm{cos}(\theta_{j,k}) + \frac{1}{3}\right)^2.
\end{equation}
The indices $j$ and $k$ run over $i$'s four nearest neighbor molecules and the angle $\theta_{j,k}$ is formed by the oxygen atoms of molecules $i$, $j$, and $k$ with $i$ in the center. The order parameter $q_\textrm{tet}$ is defined such that it is $1$ when the molecule forms a perfect tetrahedron with its four nearest neighbors and on average $0$ for an ideal gas. The RDFs and $q_\textrm{tet}$ in the QM region of the adaptive quantum--classical water systems perfectly match the results from full-quantum reference simulations. Consistent with previous work \cite{Fritsch2014}, we do not find any quantum effects for the tetrahedral order parameter $q_\textrm{tet}$. We conclude that the PI-based water structure in the QM region is undisturbed by the coupling to the CL particle reservoir.

\subsection{Dynamics}
\begin{figure}[ht!]
\centering
  \includegraphics[width=0.85\columnwidth]{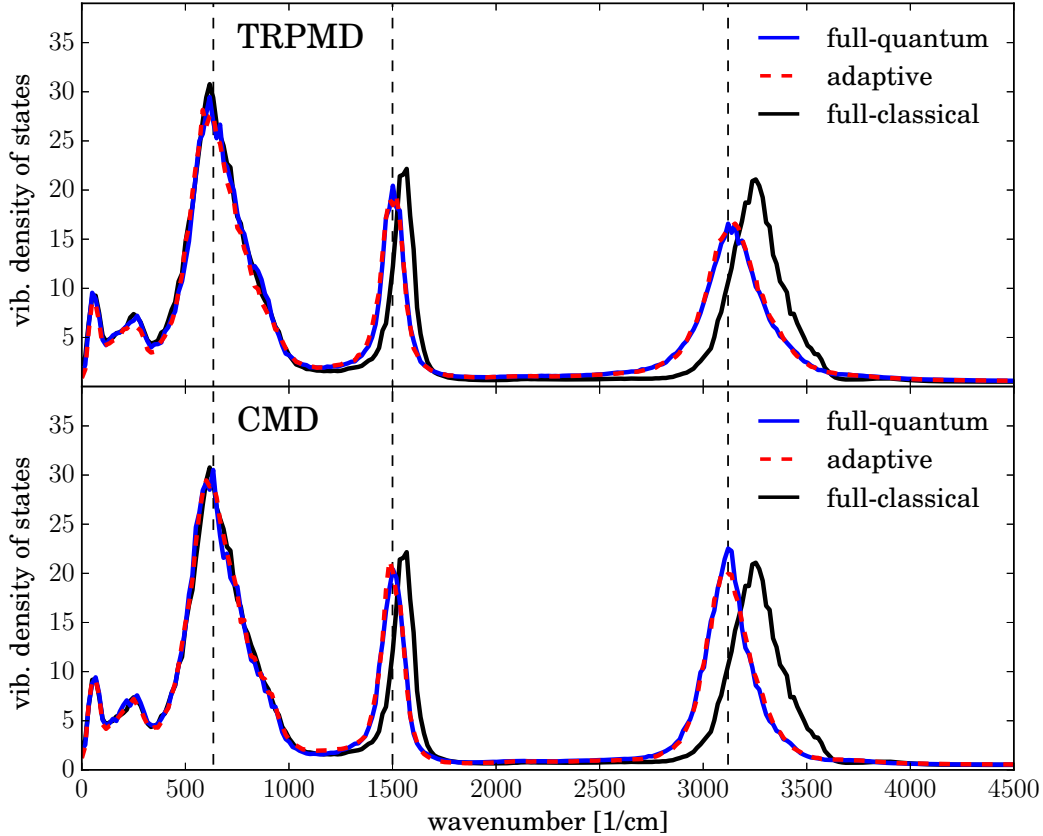}
  \caption[Vibration spectra in the QM region of quantum--classical adaptive resolution simulations of liquid water.]{Vibrational density of states in the QM region of the quantum--classical adaptive resolution simulations of liquid water and in full-quantum and full-classical reference simulations, calculated using TRPMD and CMD. The dashed vertical lines indicate the diffusion mode, the H-O-H bending mode, and the O-H stretching mode. \label{chapter7_fig5}}
\end{figure}

We also probe the dynamics in the inner QM region of the adaptive quantum--classical water systems. First, we calculate the vibrational spectrum from the water molecules' velocity autocorrelation function. We do this both via TRPMD (setup 1) and CMD (setup 2) and compare the results to full-quantum and full-classical ($P=1$) reference simulations (Fig. \ref{chapter7_fig5}). The vibrational dynamics in the QM region perfectly reproduces the full-quantum reference data, both for CMD and TRPMD. While CMD and TRPMD give similar results, the classical system shows blue shifts in the H-O-H bending and O-H stretching modes. The spectra also agree with the results from Fritsch et al. \cite{Fritsch2014}.

As hydrogen bonds play a critical role in the behavior of water \cite{Eisenberg1969,Stanley1990,Teixeira1993}, we additionally assess the hydrogen bonding kinetics of water in the QM region. The breaking and forming of hydrogen bonds can be characterized by the correlation function
\begin{equation}\label{eq:hbondfluct}
C(t) = \frac{\langle h(0)h(t) \rangle}{\langle h\rangle},
\end{equation}
which measures fluctuations in the hydrogen bond populations throughout the system \cite{Luzar1996,Luzar1996a}. The hydrogen bond population operator $h(t)$ is $1$, if a particular pair of molecules is hydrogen bonded and $0$ otherwise ($\langle h\rangle$ denotes the average of $h(t)$). We consider two molecules to be hydrogen bonded if the distance between their oxygen atoms is $<3.5\,\textrm{\AA{}}$ and the angle between the O-O axis and one of the O-H bonds is $<30^{\circ}$. Based on Eq. \ref{eq:hbondfluct}, we can determine the hydrogen bond relaxation rate $k(t)$ as
\begin{equation}\label{eq:hbondfluct2}
k(t) = -\frac{\textrm{d}C(t)}{\textrm{dt}}.
\end{equation}
The quantity $-k(t)$ can be interpreted as the average rate  of  change  of  hydrogen bonds that are broken at time $t$ later. It has been widely used in studies of the hydrogen bond kinetics in liquid water \cite{Luzar1996,Luzar1996a,Luzar2000,Luzar2000a,Benjamin2005,Winter2008}.

\begin{figure}[ht!]
\centering
  \includegraphics[width=0.75\columnwidth]{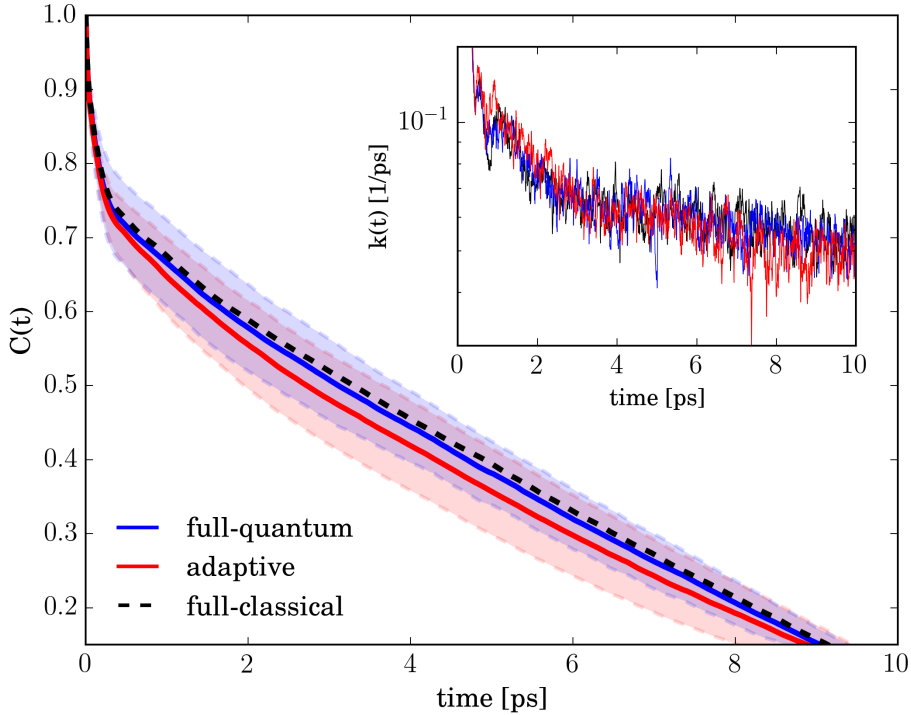}
  \caption[Hydrogen bond population fluctuations in the QM region of quantum--classical adaptive resolution simulations of liquid water.]{Hydrogen bond population fluctuations characterized via the correlation function $C(t)$ in the QM region of the quantum--classical adaptive resolution simulations and in full-quantum and full-classical reference simulations. The shaded regions indicate the standard deviations of the data corresponding to the full-quantum and adaptive simulations. Inset: average hydrogen bond relaxation rates k(t) in a semi-log plot.
\label{chapter7_fig7}}
\end{figure}

We employ the centroids for measuring the hydrogen bonds and calculate $C(t)$ via CMD using setup 2. We also perform full-quantum and full-classical reference simulations. The results are shown in Fig. \ref{chapter7_fig7}. The hydrogen bonding kinetics in the QM region of the adaptive system reproduces the full-quantum reference within the statistical error. We conclude that the hydrogen bond kinetics in the QM region of the adaptive simulations is well preserved. Furthermore, we observe no quantum effects. The classical and the quantum system behave the same in their hydrogen bonding dynamics.

We conclude that both the water structure and the PI-based dynamics in the QM region are unaffected by the coupling to the CL domain. Importantly, we have shown that one can apply both CMD and TRPMD in the proposed quantum--classical adaptive resolution simulation scheme.

\subsection{Particle fluctuations}
\begin{figure}[ht!]
\centering
  \includegraphics[width=0.8\columnwidth]{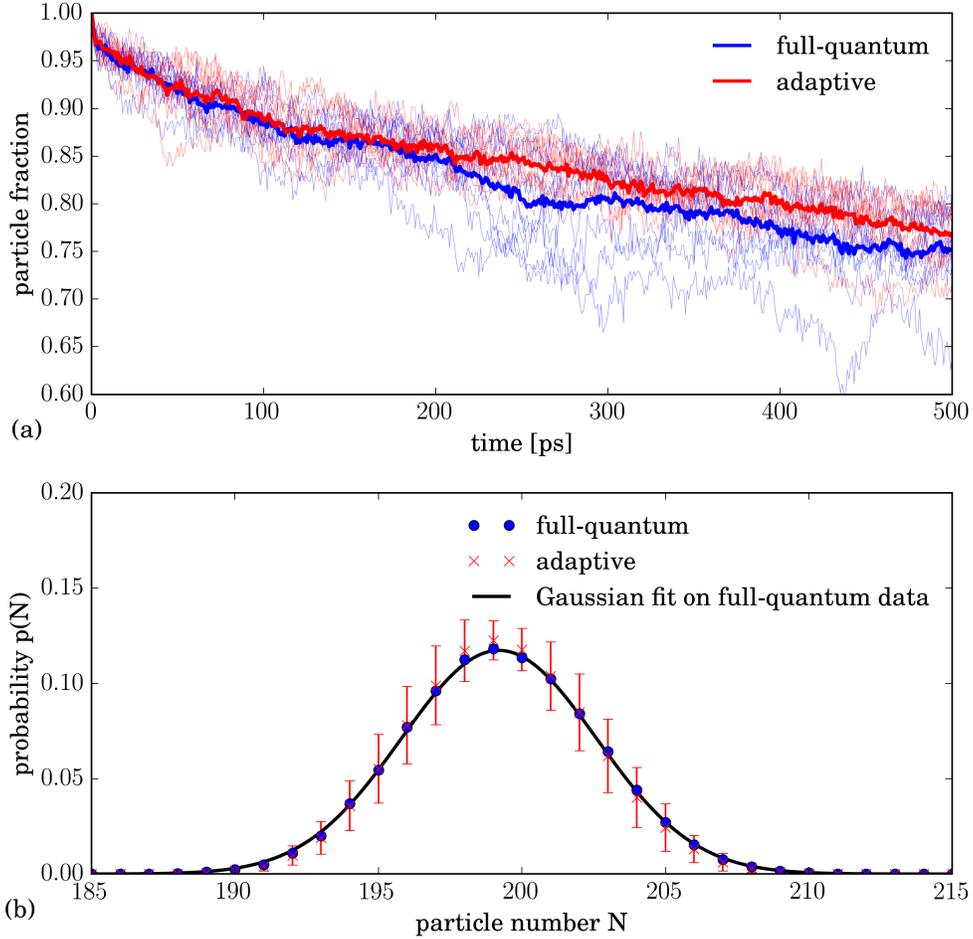}
  \caption[Particle fluctuations in the quantum--classical adaptive resolution simulations of liquid water.]{(a) Fraction of molecules that resided at the beginning of the simulation in the inner quantum region and still stay there after time $t$ for an adaptive system (setup 1) and for a full-quantum reference system, considering the same subregion. The thick lines denote the average from 10 simulations (thin lines). (b) Particle number probability distribution of the inner part of the quantum region in the adaptive simulations and of the same volume in a reference full-quantum simulation. The black curve is a Gaussian fit to the latter ($\mu = 199.2$, $\sigma = 3.4$). The error bars denote the standard deviation of the adaptive simulation data. The statistical error of the full-quantum data is similar. \label{chapter7_fig8}}
\end{figure}

It is important that the proposed method allows for a free flow of particles through the HY region without any barriers. The QM region must behave as if embedded in an overall QM environment. To test this, we label all molecules that reside at the beginning of a simulation in the inner QM region, leaving a buffer of $0.25\,\textrm{nm}$. We then track how many of the labeled particles remain in the inner QM region after time $t$, and compare this to a full-quantum reference simulation in which we label and track all molecules in a similar subregion of the system. Note that we keep the thermostat on the ring polymers' centroid modes for these simulations. The results are presented in Fig. \ref{chapter7_fig8} (a) and show that the particles diffuse out of the QM region in the adaptive setup in a similar fashion as in the full-quantum system.

Additionally, we measure the particle number fluctuations in the inner QM region (Fig. \ref{chapter7_fig8} (b)). The fluctuations match the full-quantum reference nearly perfectly. Note that the data for the adaptive system in Fig. \ref{chapter7_fig8} correspond to setup 1. All other setups show similar behavior.

The results indicate that the HY region allows a free exchange of molecules between the CL and QM regions and that the QM region exchanges particles with its environment as if embedded in a full-quantum environment. Considering the complexity of the setup, the application of a correction force in the HY region, and the different structure and thermodynamics in the CL and QM subsystems, this is non-trivial. Because of the free flow of particles and the correct particle number fluctuations in the QM region, the scheme can, for example, be used for efficient simulations of open quantum systems.

\section{Discussion and conclusions} \label{conclusions_9}
We have proposed and validated a concurrent multiscale method for Hamiltonian adaptive resolution molecular dynamics simulations using the PI formalism. The scheme is based on a position-dependent particle mass, which controls the extension and collapse of the ring polymers. In the QM region, where the particles have their real masses, the ring polymers are extended, while in the CL region, where the mass is increased, the ring polymers collapse to point-like particles. Therefore, the interaction becomes classical and the dynamics obeys classical Newtonian mechanics in the CL region. The particles freely diffuse between the two regions and change their description on the fly. The method allows a more efficient evaluation of forces and energies in the CL domain, which leads to a speedup compared to full PI simulations. Importantly, we provide criteria that quantify to what extent such an adaptive PI setup is consistent with a bottom-up PI quantization. We want to point out that this differentiates our approach from related methodologies which are based on a direct interpolation of the forces corresponding to a classical and a PI system \cite{Poma2010,Poma2011,Potestio2012,Agarwal2015,Agarwal2016}. These techniques do not allow a Hamiltonian description of the system \cite{DelleSite2007}, which, however, is the basis for a bottom-up PI treatment in the first place. Our scheme aims at overcoming this limitation. It allows both adaptive PIMD simulations sampling quantum statistical averages as well as quantum--classical RPMD and CMD, which enable us to calculate approximate quantum dynamical quantities and time correlation functions. Finally, the method allows one not only to selectively turn on and off nuclear quantum effects in different regions but also to change the intermolecular interaction potential. In this way, one can use a more efficient, possibly coarse-grained model in the CL region. This would be useful, for example, when the CL domain only serves as a particle reservoir.

To implement our methodology in a molecular dynamics framework, a kinetic energy term needs to be introduced into the configurational energy obtained from PI quantization. At this point, our approach can be implemented in two different ways: The kinetic masses in this kinetic energy term can be chosen to be either constant throughout the whole system (CKM) or they can vary in a way similar to the particle masses that control the springs between the PI beads (AKM). The CKM approach results in a simpler scheme that can be integrated with a standard velocity Verlet integrator. However, it leads to strong thermodynamic imbalances between the CL and the QM regions and requires very small time steps due to the accelerated vibrations of the ring polymers in the CL region. On the other hand, the AKM method requires a more sophisticated integration scheme due to the position-dependent kinetic masses. We derived an integrator which is tailored to the problem, employs multiple time-stepping, and allows a symplectic integration of the equations of motion. This scheme also facilitates time steps which are much larger than in the CKM protocol and which are similar to those used in normal PI simulations. The AKM method also enables a smoother connection of the CL and QM systems, requiring a milder correction force in the HY coupling region.

The new integrator may appear complicated, but it requires little additional overhead in practice. In molecular dynamics simulations, most time is typically spent for non-bonded force calculations and for inter-processor communication. However, these two tasks do not need to be performed more often than in a standard velocity Verlet integrator. 
In its essence, our methodology elegantly decouples the change of the particles' quantum character, which is connected only to the higher modes and requires an additional decomposition step in the inner loop of the integrator, from the interatomic and intermolecular interactions, which are related to the more expensive bonded and non-bonded force calculations. 
In fact, it is only the masses of the higher modes that are position-dependent, while the masses of the centroid modes are constant.
In our implementation the integration is performed in normal mode space. Only before the calculation of the bonded and non-bonded forces, the particles' real positions are updated and the force calculation is performed in real space. Afterwards, the forces are transformed back to normal modes.
Therefore, the innermost loop of our integrator, in which the additional decomposition step occurs, does not require any inter-processor communication. When applying the methodology on systems in which only a very small part of the simulation domain is modeled quantum mechanically the additional overhead will be negligible compared to the gain in computational efficiency over a similar full QM system. We did not perform a detailed study of the speedup, though, as this depends in practice on a large number of factors, such as the system at hand, the scheme's implementation, the parallelization methodology, and the load balancing protocol (in highly parallelized simulations that employ many CPUs a suitable load balancing method that allows to concentrate computational resources in the QM region is crucial). Nonetheless, we have shown already in our previous paper that the speedup can be significant \cite{Kreis2016}. Provided the QM subsystem is small, the interatomic force computations can be accelerated by a factor of $10$ or more. Note that the presented adaptive resolution method can of course also be used in setups with different geometrical arrangements of the QM and CL regions compared to the one in this article. A typical example would be a small spherical QM domain positioned at an area of particular interest within a large CL system.

The proposed adaptive PI simulation scheme gains its efficiency by restricting the QM region to a small but relevant region in space and treating the rest of the system with a more efficient classical model. This is in contrast to other approaches that aim to alleviate the computational cost of PI simulations by modifying the PI calculations themselves, such as RPC \cite{Markland2008, Markland2008a}, higher order Trotter factorizations \cite{Takahashi1984}, or advanced thermostating techniques \cite{Ceriotti2009,Ceriotti2010,Ceriotti2011,Ceriotti2012}. Our method is complementary to these approaches and could be combined with them. For example, one could apply RPC to further reduce the numerical effort in the QM region or use a colored noise instead of a white-noise Langevin thermostat, which would allow one to employ less Trotter beads. One could also further improve the multiple time-stepping and tailor it to the investigated systems.

The applications of the proposed methodology are diverse. The scheme is useful whenever only a small subdomain of an overall large system needs to be described including PIs. This can be the case, for example, in biomolecular systems, in which the study of nuclear quantum effects has gained significant interest \cite{Lowdin1963,Rein1964,Marx1999,Schmitt1999,Haines2001,Tuckerman2002,Jeuken2007,Wu2008,Berkelbach2009,Perez2010,Marx2010,Smirnov2011,Pamuk2012,Jacquemin2014,Wang2014}. Biological systems are often complicated and quantum delocalization plays an important role usually only in a small part of the system, such as the active site of proteins \cite{Wang2014}. Our multiscale method could be used to describe the active site quantum mechanically and an efficient classical model could be employed for the rest of the system, in the same spirit of QM/MM approaches but at a different level of ``quantumness''. This would allow an extension of the accessible length and time scales compared to full path integral simulations. Similar applications of the scheme are simulations of interfaces or membranes. The possibility to selectively switch on and off the nuclear quantum effects in different regions also allows one to investigate the locality of quantum properties. One can ask, for example, how much quantum mechanically modeled environment is required to support the quantum mechanical features in a certain subregion \cite{Lambeth2010,Fritsch2012a,Fogarty2014,Agarwal2017}. This would not be possible in bulk PI simulations in a straightforward way. Furthermore, the method enables an efficient simulation of a quantum grand canonical ensemble: a QM region can be coupled to a large particle reservoir, which itself is described classically. Yet another interesting possible application of our methodology is its combination with the aforementioned QM/MM techniques, which concurrently couple \textit{ab initio} and classical empirical force fields. Recently, Boereboom et al. \cite{Boereboom2016} proposed an adaptive QM/MM method based on the Hamiltonian adaptive resolution scheme. The latter is also used in the PI-based adaptive resolution scheme presented in this article. In fact, although the interatomic potentials employed in this work are empirical force fields, the forces and energies could also come from \textit{ab initio} calculations. Therefore, one could combine our approach with the one from Boereboom et al. and construct a Hamiltonian adaptive QM/MM scheme that also incorporates a multiscale treatment of PIs. 

Finally, we would like to point out that the derived concurrent multiscale PI simulation methodology has been implemented in the ESPResSo++ package \cite{Halverson2013} and is publicly available.

\begin{acknowledgments}
K. Kreis is recipient of a fellowship funded through the Excellence Initiative (DFG/GSC 266). M.E.T. acknowledges support from the National Science Foundation, grant CHE-1566085. The authors thank Joe Rudzinski for a careful reading of the manuscript and his helpful suggestions.
\end{acknowledgments}

%

\end{document}